\documentclass[onecolumn,authoryear]{els-mrws} 

\usepackage{amsmath,amssymb,amsfonts,amsthm,makeidx,graphicx}
\usepackage{txfonts}
\usepackage{helvet}

\usepackage{hyperref}  
\usepackage[capitalise,nameinlink,noabbrev]{cleveref}  
\usepackage{lineno}
\usepackage{booktabs}

\def\urltilda{\kern -.15em\lower .7ex\hbox{\~{}}\kern .04em}
\def\urldot{\kern -.10em.\kern -.10em}
\def\urlhttp{http\kern -.10em\lower -.1ex\hbox{:}\kern -.12em\lower 0ex\hbox{/}\kern -.18em\lower 0ex\hbox{/}}

\def\sp{\kern 0.08333em}

\newcommand{\muegamma}{$\mu^{+} \rightarrow e^{+} \gamma$}

\newcommand{\mueee}{$\mu^{+} \rightarrow e^{+}e^{+}e^{-}$}

\newcommand{\mue}{$\mu^{-}\to e^{-}$}
\newcommand{\mup}{$\mu^{-}\to e^{+}$}

\newcommand{\Deg}{\(^{\circ} \) \kern-.4em \ }

\IfFileExists{upgreek.sty}%
  { \usepackage{upgreek}%
  }%
  { \IfFileExists{textgreek.sty}%
    { \usepackage{textgreek}%
    }%
    {%
    }%
  }%

\newcommand{\captionfonts}{\small}

\makeatletter  
\long\def\@makecaption#1#2{%
  \vskip\abovecaptionskip
  \sbox\@tempboxa{{\captionfonts #1: #2}}%
  \ifdim \wd\@tempboxa >\hsize
    {\captionfonts #1: #2\par}
  \else
    \hbox to\hsize{\hfil\box\@tempboxa\hfil}%
  \fi
  \vskip\belowcaptionskip}
\makeatother   

\begin{document}

\chapter{Experimental Searches for Muon to Electron Conversion in Nuclei}\label{chap1}

\author[1]{Yoshitaka Kuno}%

\address[1]{\orgname{The University of Osaka}, \orgdiv{Research Center for Nuclear Physics}, \orgaddress{1-1 Machikaneyama, Toyonaka, Osaka, Japan 560-0043}}

\maketitle

\begin{abstract}[Abstract]
Charged lepton flavor violation (CLFV) provides a highly sensitive probe of physics beyond the Standard Model. This experimental review discusses \mue\ conversion in nuclei, which is a muon CLFV process characterized by a clean monoenergetic-electron signature and the ability to exploit large samples of stopped muons. It describes the physics motivation, basic experimental principles, key background sources, and search strategies, as well as the status and prospects of past and next-generation experiments such as COMET and Mu2e. Future experimental concepts may extend the sensitivity even further. If a signal is observed, comparisons among results obtained with different target nuclei and with those from complementary searches for other muon CLFV decay modes will be essential for identifying the underlying interaction.
\end{abstract}

\section{Introduction}

\subsection{Charged Lepton Flavor Violation as a Probe of Physics Beyond the Standard Model}

Before the discovery of neutrino oscillations, neutrinos were assumed to be massless in the Standard Model, and lepton flavor was conserved separately in each generation. The observation of neutrino oscillations fundamentally changed this picture: neutrinos have nonzero masses and mix with one another. Lepton flavor is therefore not an exact symmetry of nature.

In the charged-lepton sector, however, no flavor-violating process has yet been observed. Processes such as \muegamma\ decay, \mueee\ decay, $\mu^- N \to e^- N$ conversion in nuclei, and rare $\tau$ decays such as $\tau \to \ell \gamma~(\ell = e, \mu)$ are known as charged lepton flavor violating (CLFV) processes. Here, $N$ denotes a nucleus, and the term \mue\ conversion will be used throughout this review to refer to the process $\mu^- N \to e^- N$. The non-observation of these processes, together with the extremely small rates predicted in the Standard Model with massive neutrinos, makes CLFV searches among the most sensitive probes of physics beyond the Standard Model. For broader reviews of CLFV theory and experimental searches, see 
\citet{Kuno:1999jp,Marciano:2008zz,deGouvea:2013zba,Mihara:2013zna,Bernstein:2013hba,Calibbi:2017uvl}.

Within the framework of effective field theory, muon CLFV processes provide exceptionally sensitive probes of new physics. For dimension-six operators with couplings of order $\mathcal{O}(1)$, present limits can already correspond to effective new-physics scales as high as $\mathcal{O}(10^{3}$\text{--}$10^{4})\,\mathrm{TeV}$~\citep{EuropeanStrategyforParticlePhysicsPreparatoryGroup:2019qin,Davidson:2022jai}. Furthermore, next-generation muon CLFV searches aim to improve experimental sensitivities by up to four orders of magnitude. Because the rates induced by these operators typically scale as $\Lambda^{-4}$, with $\Lambda$ denoting the effective new-physics scale, this improvement would extend the reach in new-physics scale by roughly one order of magnitude. Such sensitivity is among the highest attainable in rare-process searches, and the corresponding scales far exceed those directly accessible at present or planned high-energy colliders.

Many extensions of the Standard Model can enhance CLFV rates to levels accessible to current or next-generation experiments. Examples include supersymmetric models, seesaw scenarios, extended Higgs sectors, leptoquark models, and theories with additional gauge interactions. Since different CLFV processes are sensitive to different effective operators, comparisons among \muegamma\ decay, \mueee\ decay, and \mue\ conversion can provide important information on the structure of possible new physics.

In recent years, the development of high-intensity muon sources has enabled the next generation of muon CLFV searches with substantially improved sensitivity. The current limits and expected future reach for representative muon CLFV processes are summarized in \Cref{tab:clfv-current-future}. Among them, \mue\ conversion in nuclei is especially promising. It benefits from a distinctive experimental signature and probes a broad class of interactions involving both leptons and quarks.

\begin{table}[tb]
\setlength{\heavyrulewidth}{1pt}
\centering
\caption{Current 90\% C.L. limits and expected future sensitivities for three major muon CLFV processes.}
\vspace{3mm}
\label{tab:clfv-current-future}
\begin{tabular}{lll} 
\toprule
Process & Current limit &  Expected future sensitivity  \\
\midrule
\muegamma\ & $<1.5 \times 10^{-13}$ (MEG II~\citep{MEGII:2025gzr}) & $\sim 6 \times 10^{-14}$ (MEG II~\citep{MEGII:2018kmf}) \\
\mueee\ & $<1.0\times 10^{-12}$ (SINDRUM~\citep{BELLGARDT19881})  & $\sim 10^{-16}$ (Mu3e~\citep{Mu3e:2020gyw})\\
$\mu^{-}N \to e^{-}N$ & $<7\times 10^{-13}$ (Au, SINDRUM II~\citep{Bertl:2006te}) & $\sim 10^{-16}$ (Al, COMET~\citep{Kuno:2013mha}) \\
& & $\sim 10^{-16}$ (Al, Mu2e~\citep{bartoszek2015mu2etechnicaldesignreport}) \\
\bottomrule
\end{tabular}
\end{table}

In the following, we first review the Standard Model expectation for CLFV and then discuss the role of \mue\ conversion in comparison with other major muon CLFV processes.


\subsection{Standard Model Expectation for CLFV}

\begin{figure}[tbh!]
\centering
	 \includegraphics[width=0.6\textwidth]{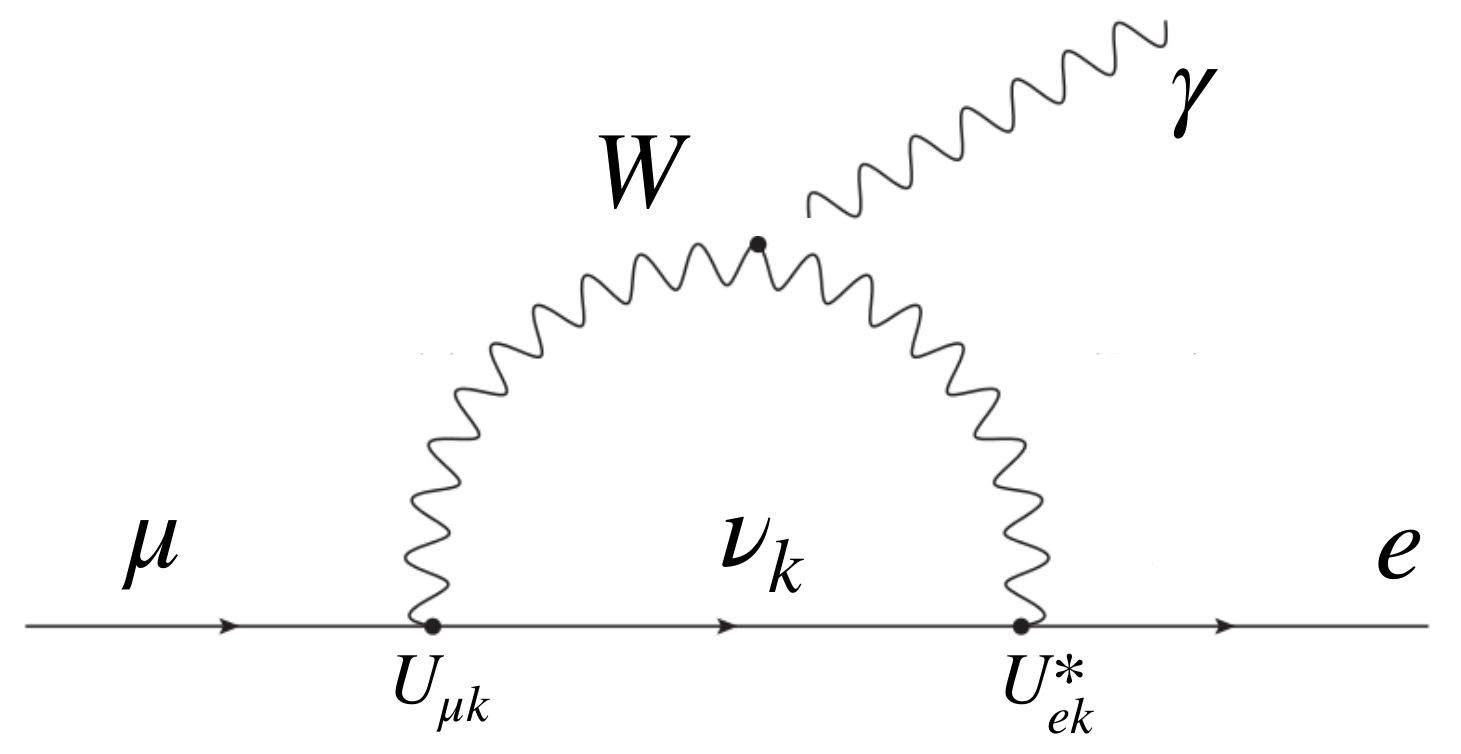}
  \caption{Representative Feynman diagram for the \muegamma\ decay process in the Standard Model with massive neutrinos.}
  \label{fig:muegamma-feynman}
\end{figure}

CLFV processes arise in the Standard Model with massive neutrinos through loop diagrams involving neutrino mixing. For the rare \muegamma\ decay, a representative diagram is shown in \Cref{fig:muegamma-feynman}. In the Standard Model with massive neutrinos, the predicted branching ratio, denoted by ${\mathrm{BR}}_{\mathrm{SM}}$, is
\begin{align}
{\mathrm{BR}}_{\mathrm{SM}}(\mu\to e\gamma) &=
\frac{3\alpha}{32\pi}
\left|
\sum_{k=1,2,3}
U_{\mu k}U^{*}_{e k}
\frac{m_{\nu_k}^2}{M_W^2}
\right|^2 \nonumber \\
&=
\frac{3\alpha}{32\pi}
\left|
U_{\mu 3}U^{*}_{e 3}
\frac{\Delta m_{31}^2}{M_W^2}
+
U_{\mu 2}U^{*}_{e 2}
\frac{\Delta m_{21}^2}{M_W^2}
\right|^2 .
\label{eq:sm-muegamma}
\end{align}
Here, $\Delta m_{ij}^2=m_{\nu_i}^2-m_{\nu_j}^2$ for $i,j=1,2,3$, where $m_{\nu_i}$ denotes the mass of the $i$th neutrino mass eigenstate. $M_W$ is the mass of the $W$ boson, $\alpha$ is the fine-structure constant, and $U_{\ell k}$ are elements of the Pontecorvo--Maki--Nakagawa--Sakata (PMNS) lepton mixing matrix, with $\ell =e,\mu,\tau$ and $k=1,2,3$. The unitarity of the PMNS matrix leads to a lepton-sector analogue of the Glashow--Iliopoulos--Maiani (GIM) mechanism. Consequently, the amplitude is suppressed by factors of order $\Delta m_{ij}^2/M_W^2$, while the branching ratio is suppressed by the squares of these factors. Since neutrino mass splittings are tiny compared with the electroweak scale, ${\mathrm{BR}}_{\mathrm{SM}}(\mu\to e\gamma)\sim\mathcal{O}(10^{-54})$, which is negligibly small~\citep{Petcov:1976ff,Bilenky:1977du,Cheng:1976uq,Marciano:1977wx,Lee:1977qz,Lee:1977tib}. Similar suppressions occur for other CLFV modes, such as \mueee\ decay~\citep{Petcov:1976ff,Hernandez-Tome:2018fbq}. Therefore, any observation of CLFV would constitute clear evidence for new physics beyond the Standard Model.

\subsection{Muon CLFV Processes and the Role of \texorpdfstring{\mue}{mu-e}\ Conversion}
\label{sec:threeCLFVprocesses}

Muon-related searches provide some of the most sensitive probes of CLFV, owing to the availability of intense muon sources and the clean experimental signatures of rare muon processes. The muon CLFV processes discussed above are complementary, since they test the same underlying CLFV physics in different ways and are sensitive to different operators.

The experimental signatures differ significantly among the three muon CLFV processes. Searches for the \muegamma\ and \mueee\ processes use positive muons decaying at rest. The \muegamma\ decay is a two-body decay characterized by a positron and a photon emitted back-to-back, each carrying an energy close to $m_\mu/2$. \mueee\ decay has a fully charged final state with constrained kinematics, in which the three charged particles originate from a common vertex, their total reconstructed energy is consistent with $m_\mu$, and the vector sum of their momenta is consistent with zero. In contrast, \mue\ conversion occurs in a muonic atom and produces a monoenergetic electron while the nucleus remains the same. A \mue\ conversion search can therefore be formulated primarily as a narrow momentum-peak measurement, rather than as a coincidence measurement involving multi-particle reconstruction as in \muegamma\ and \mueee\ searches.

\begin{figure}[tbh!]
\centering
	 \includegraphics[width=0.6\textwidth]{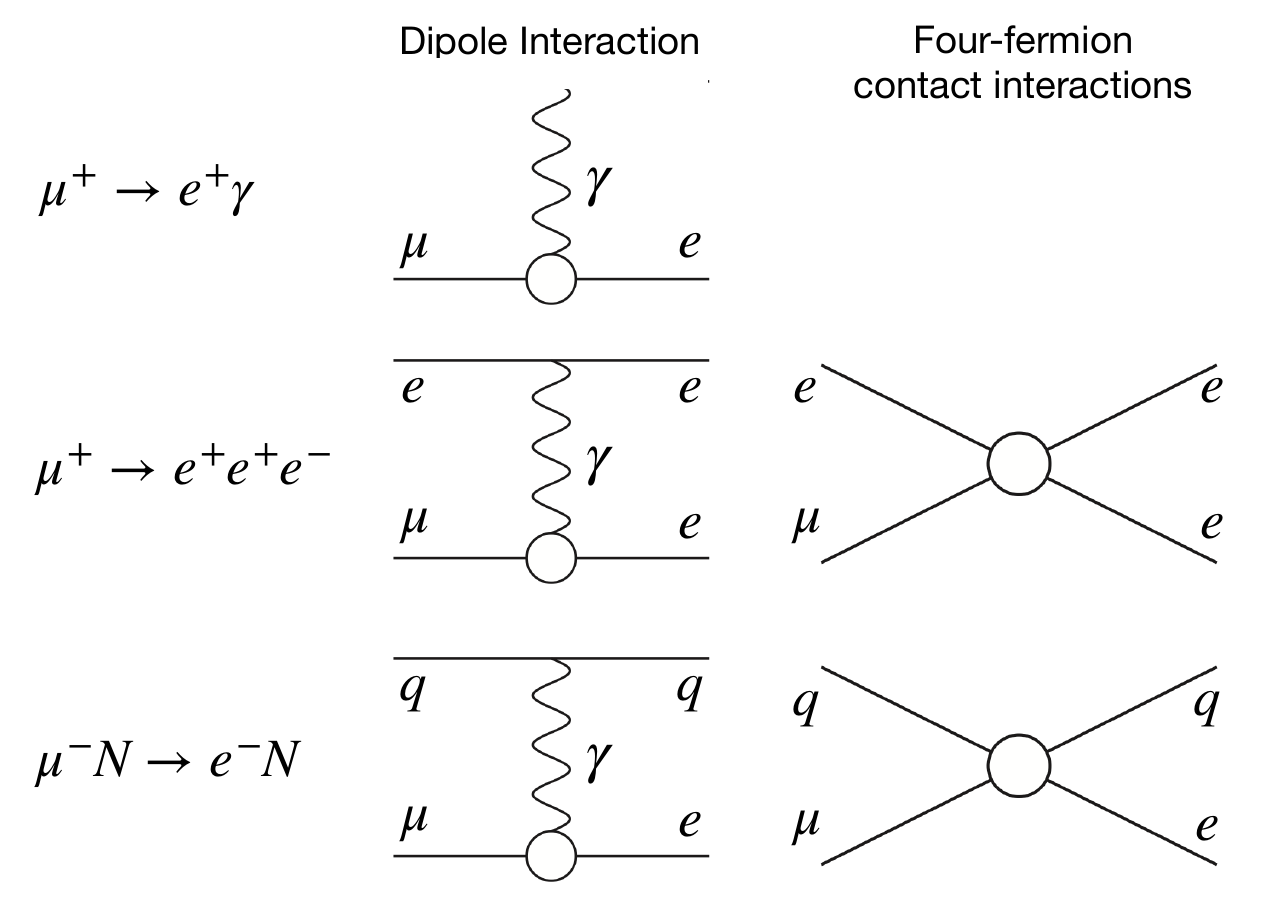}
\caption{Schematic illustration of dipole and four-fermion contact contributions to representative muon CLFV processes. The radiative decay \muegamma\ directly probes a dipole interaction, whereas \mueee\ and $\mu^{-}N \to e^{-}N$ can receive contributions from both dipole and contact interactions. The contact contribution to \mueee\ arises from four-lepton operators, while that to $\mu^{-}N \to e^{-}N$ arises from two-lepton--two-quark operators.}
  \label{fig:clfv-topologies}
\end{figure}

The three processes also probe different classes of effective interactions, as summarized in \Cref{fig:clfv-topologies}. The radiative decay \muegamma\ is especially sensitive to dipole operators, whereas \mueee\ probes both dipole interactions and four-lepton contact interactions. The \mue\ conversion process is sensitive not only to dipole interactions but also to four-fermion contact interactions involving two leptons and two quarks. It therefore provides direct access to CLFV interactions involving quarks and nuclei. In this sense, \mue\ conversion uses the nucleus as a laboratory for testing CLFV physics.

For these reasons, \mue\ conversion occupies a central role in the current muon CLFV program. Its single-particle signature and broad operator sensitivity make it one of the most powerful processes for discovering or constraining physics beyond the Standard Model.

\subsection{Scope of This Review}

This review summarizes searches for \mue\ conversion in nuclei, with emphasis on their physics motivation, experimental principles, and techniques used in past, current, and next-generation experiments. The theoretical framework, including the relevant effective interactions and nuclear and atomic effects, is presented only as needed to define the process, explain its sensitivity to CLFV, and interpret experimental results. We discuss the signal-electron signature, principal backgrounds, beam and detector requirements, current limits, and future prospects. Rather than describing individual detector subsystems in detail, the review focuses on the experimental concepts that make \mue\ conversion a sensitive probe of new physics.

%
%
\section{Phenomenology of \texorpdfstring{\mue}{mu-e}\ Conversion in Nuclei}

\subsection{Muonic Atoms and \texorpdfstring{\mue}{mu-e}\ Conversion in Nuclei}

Experimental searches for \mue\ conversion in nuclei use muonic atoms, in which a negatively charged muon, $\mu^-$, is bound to a nucleus. When a $\mu^{-}$ stops in a target material, it is initially captured by the Coulomb field of the nucleus into an orbit with high principal quantum number and high angular momentum. The muon then rapidly cascades down to the $1\mathrm{s}$ ground state of the muonic atom through electromagnetic transitions, emitting muonic X-rays and/or Auger electrons. 

Once in the $1\mathrm{s}$ ground state, the bound muon remains in the atom until it either decays in orbit or is captured by the nucleus through ordinary muon capture. If CLFV occurs, the bound muon can instead undergo neutrinoless conversion into an electron,
\begin{equation}
\mu^{-} + N(A,Z) \to e^{-} + N(A,Z),
\end{equation}
where $N(A,Z)$ denotes a nucleus with mass number $A$ and atomic number $Z$. This process is referred to as \mue\ conversion in nuclei.  It changes the muon and electron flavor numbers, $L_\mu$ and $L_e$, by $\Delta L_{\mu}=-1$ and $\Delta L_e=+1$, respectively, while conserving the total lepton number.

The final-state nucleus may either remain in its ground state, corresponding to elastic \mue\ conversion, or be left in an excited state. When elastic \mue\ conversion is induced by spin-independent CLFV interactions, the contributions from individual nucleons can add coherently at the amplitude level, substantially enhancing the \mue\ conversion rate. This coherently enhanced, spin-independent ground-state contribution is conventionally referred to as coherent \mue\ conversion.

Spin-dependent CLFV interactions can also induce elastic conversion to the
nuclear ground state, but only for target nuclei with nonzero spin. Such
contributions are not coherently enhanced and depend instead on nuclear spin
matrix elements. Even for such nuclei, spin-independent interactions can still 
generate a coherently enhanced elastic ground-state transition.

Transitions to excited nuclear states are referred to as inelastic \mue\
conversion. These transitions are not coherently enhanced and depend
sensitively on the relevant nuclear transition matrix elements. The
spin-independent and spin-dependent CLFV operators responsible for these
contributions will be discussed in detail later in \Cref{sec:eft}.

\subsection{Experimental Signature of \texorpdfstring{\mue}{mu-e}\ Conversion}
\label{sec:eventsignature}

A characteristic feature of the \mue\ conversion process in a muonic atom is that a single monoenergetic electron with a well-defined energy is emitted as a consequence of energy conservation. From the two-body kinematics of the final state, the electron energy in elastic \mue\ conversion, $E_{\mu e}$, is given by
\begin{align}
E_{\mu e} &= \frac{(m_{N} + m_{\mu} - B_{\mu})^2 - m_{N}^2 + m_{e}^2}{2(m_N + m_{\mu} - B_{\mu})} \nonumber \\
&\approx  m_{\mu} - B_{\mu} - E_{\rm recoil}.
\label{eq:muesignalenergy}
\end{align}
Here, $m_{\mu}$ and $m_e$ are the muon and electron masses, respectively,
$m_N$ is the mass of the nucleus $N(A,Z)$, and $B_{\mu}$ is the binding
energy of the $1\mathrm{s}$ state of the muonic atom. 
In the second expression, the electron mass has been neglected, and the
nuclear recoil energy after \mue\ conversion has been approximated as
$E_{\rm recoil} \approx (m_{\mu}-B_{\mu})^2/(2m_N)$.\footnote{For inelastic \mue\ conversion to an excited nuclear state, the signal-electron energy, $E_{\mu e}^{\ast}$, is
given by
\begin{equation}
E_{\mu e}^{\ast}
=
\frac{
(m_{N}+m_{\mu}-B_{\mu})^2
-
(m_{N}+E_{\rm exc})^2
+
m_e^2
}{
2(m_N+m_{\mu}-B_{\mu})
}
\approx
m_{\mu}-B_{\mu}-E_{\rm recoil}^{\ast}-E_{\rm exc},
\label{eq:inelastic}
\end{equation}
where $E_{\rm exc}$ is the excitation energy of the final nuclear state, and $E_{\rm recoil}^{\ast}$ is the corresponding nuclear recoil energy.}

\begin{figure}[tbh!]
\centering
	 \includegraphics[width=0.6\textwidth]{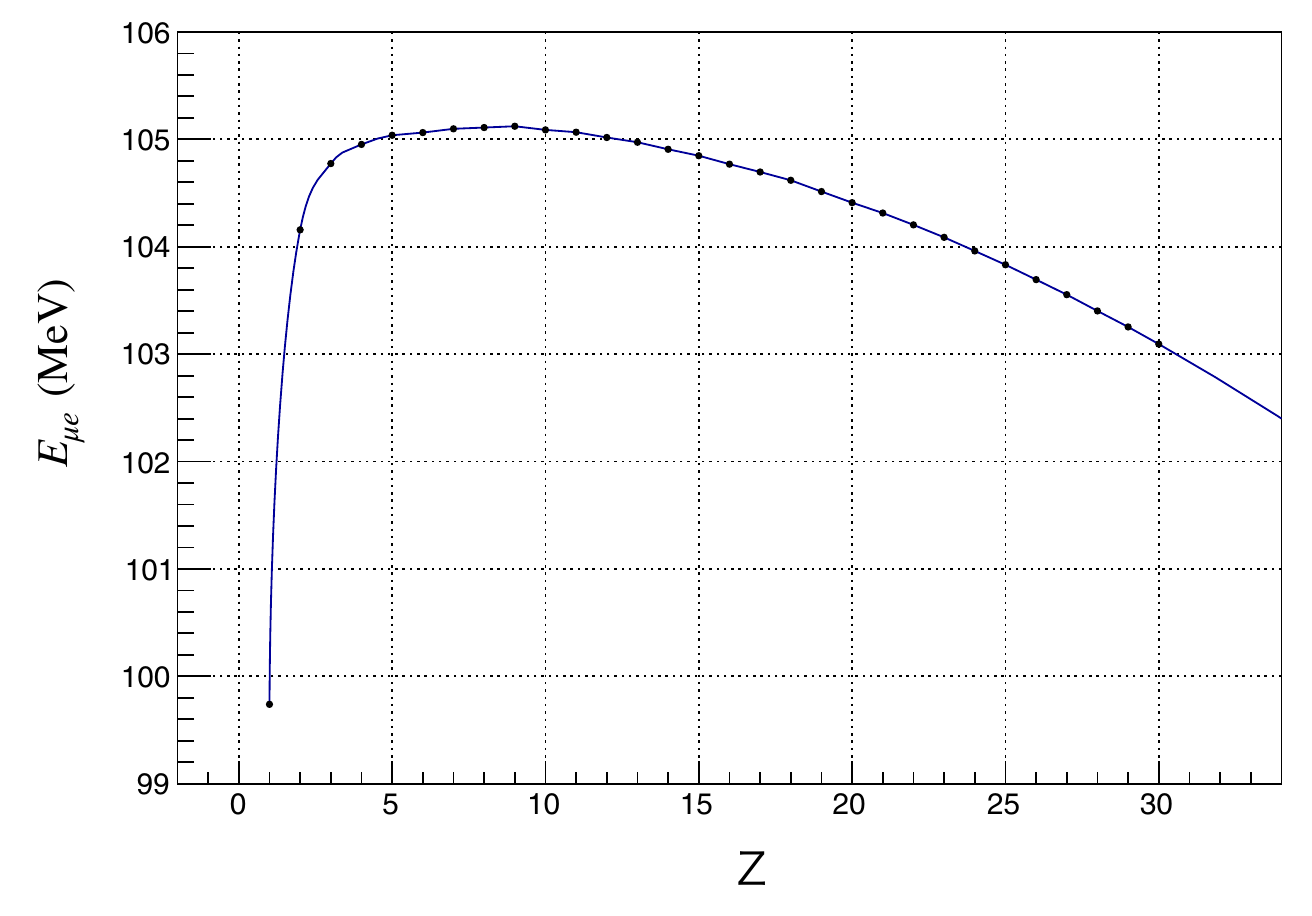}
 \caption{Signal-electron energy $E_{\mu e}$ for elastic \mue\ conversion to the nuclear ground state as a function of the atomic number $Z$ for low-$Z$ target nuclei. Adapted from \citet{10.1093/ptep/ptz125}, licensed under CC BY 4.0.}
  \label{fig:emuz}
\end{figure}

The signal-electron energy, $E_{\mu e}$, depends on the target nucleus. This dependence arises from the binding energy of the muon in the $1\mathrm{s}$ state of the muonic atom, $B_\mu$, and from the nuclear recoil energy, $E_{\rm recoil}$. In the hydrogen-like approximation, $B_\mu \simeq (1/2)m_{\mu N}(Z\alpha)^2$, where $m_{\mu N}$ is the reduced mass of the muon--nucleus bound system.
For example, $E_{\mu e}=104.97~\mathrm{MeV}$ for aluminum, $104.3~\mathrm{MeV}$ for titanium, and $94.9~\mathrm{MeV}$ for lead. \Cref{fig:emuz} shows the signal-electron energy for elastic \mue\ conversion as a function of atomic number $Z$ for low-$Z$ nuclei.

Radiative corrections to the electron spectrum of the \mue\ conversion signal have also been calculated~\citep{Szafron:2016cbv,Fontes:2024yvw}. Vacuum-polarization effects shift the position of the signal peak slightly downward, while soft-photon emission produces a low-energy tail below the nominal signal-electron energy. The vacuum-polarization shift is about $-2.7~\mathrm{keV}$ for aluminum and of order $-10~\mathrm{keV}$ for titanium, with its magnitude increasing to $\mathcal{O}(100~\mathrm{keV})$ for high-$Z$ nuclei.\footnote{For a pointlike nucleus, using a nonrelativistic muon wave function and assuming $m_e \ll m_{\mu}$,  the leading vacuum-polarization correction to the bound-muon energy, $\Delta E_{\mathrm{VP}}$, is given by~\citep{Szafron:2016cbv}
\begin{equation}
\Delta E_{\mathrm{VP}}
=
\frac{\alpha}{\pi} (Z\alpha)^2 m_{\mu}
\left[
\frac{11}{9}
-\frac{2}{3}\log\left(\frac{2m_\mu Z\alpha}{m_e}\right)
\right].
\label{eq:vacuumpolarization}
\end{equation}
For the target nuclei relevant to \mue\ conversion searches, $\Delta E_{\mathrm{VP}}<0$. Vacuum polarization therefore makes the bound-muon energy more negative.
}

\subsection{Conversion Ratio for \texorpdfstring{\mue}{mu-e}\ Conversion}

The conversion ratio (CR)\footnote{This quantity is often referred to as a branching ratio, but the term conversion ratio is used in this review for clarity.} for the \mue\ conversion process is defined as the \mue\ conversion rate normalized to the total ordinary nuclear muon-capture rate. It is given by
\begin{equation}
\mathrm{CR}(\mu^{-}N \rightarrow e^{-}N)
\equiv
\frac{\Gamma(\mu^{-}N \rightarrow e^{-}N)}
{\Gamma_{\rm cap}} ,
\label{eq:conversion-rate}
\end{equation}
where $\Gamma(\mu^{-}N \rightarrow e^{-}N)$ denotes the \mue\ conversion rate and $\Gamma_{\rm cap}$ represents the total ordinary nuclear muon-capture rate in the same nucleus. This definition has been the standard convention since the earliest experimental searches~\citep{PhysRev.100.1490}.\footnote{An alternative normalization, in which the \mue\ conversion rate is normalized to the free-muon decay rate, has also been proposed~\citep{Borrel:2024ylg}.} Unless otherwise specified, experimental sensitivities quoted in this review refer to the \mue\ conversion ratio.

\subsection{Time Distribution of the \texorpdfstring{\mue}{mu-e}\ Conversion Signal}
\label{sec:timedistribution}

The time distribution of the \mue\ conversion signal follows the disappearance rate of the muonic atom, $\Gamma_{\mu\text{-atom}}$, which is governed by muon decay in orbit and muon capture in nuclei. This disappearance rate, equal to the inverse muonic-atom lifetime $\tau_{\mu\text{-atom}}^{-1}$, is given by
\begin{equation}
\Gamma_{\mu\text{-atom}} = \frac{1}{\tau_{\mu\text{-atom}}}
=
Q\Gamma_{\rm free}
+
\Gamma_{\rm cap} .
\end{equation}
The first term, $Q\Gamma_{\rm free}$, describes muon decay in orbit, where $\Gamma_{\rm free}$ is the free-muon decay rate. The Huff factor $Q$ accounts for the reduction of the decay rate of a bound muon relative to that of a free muon. This reduction arises mainly from bound-state effects, including the reduced phase space available to the decay electron due to the muon binding energy. For example, $Q \simeq 0.975$ for iron ($Z=26$), while $Q \simeq 0.844$ for lead ($Z=82$)~\citep{HUFF1961288,PhysRevC.35.2212,Uesaka:2026tdd}. Bound-state effects also modify the electron energy spectrum of muon decay in orbit (DIO), especially for nuclei with high atomic number $Z$. Since DIO is an important background source for \mue\ conversion searches, it is discussed in more detail in \Cref{sec:dio}.

The second term, $\Gamma_{\rm cap}$, describes ordinary muon capture in nuclei. This rate increases strongly with the atomic number of the target nucleus. For low-$Z$ nuclei, the $1\mathrm{s}$ bound-muon wave function is approximately hydrogen-like, and its density near the origin scales as $Z^3$. Combined with the number of protons available for capture, which scales as $Z$, this gives the familiar approximate $Z^4$ scaling of the muon-capture rate.

For realistic nuclei, however, the muon-capture rate is controlled not simply by the nuclear charge $Z$, but by the overlap between the bound-muon wave function and the nuclear distribution. In particular, for high-$Z$ nuclei, the muonic Bohr radius becomes comparable to or smaller than the nuclear radius, and finite-nuclear-size effects modify the bound-muon wave function in the nuclear region. As a result, the effective overlap does not increase as rapidly as expected from the simple point-nucleus $Z^4$ scaling. This behavior is often parameterized in terms of the effective atomic number $Z_{\rm eff}$, which incorporates finite-nuclear-size and bound-state effects~\citep{K.W.-Ford:1968aa}.  In this approximation, the muon-capture rate scales roughly as $\Gamma_{\rm cap} \sim Z_{\rm eff}^{4}$. Nevertheless, the muon-capture rate generally increases with $Z$, so the muonic-atom lifetime becomes shorter for higher-$Z$ nuclei.

\begin{figure}[tb]
\centering
	 \includegraphics[width=0.8\textwidth]{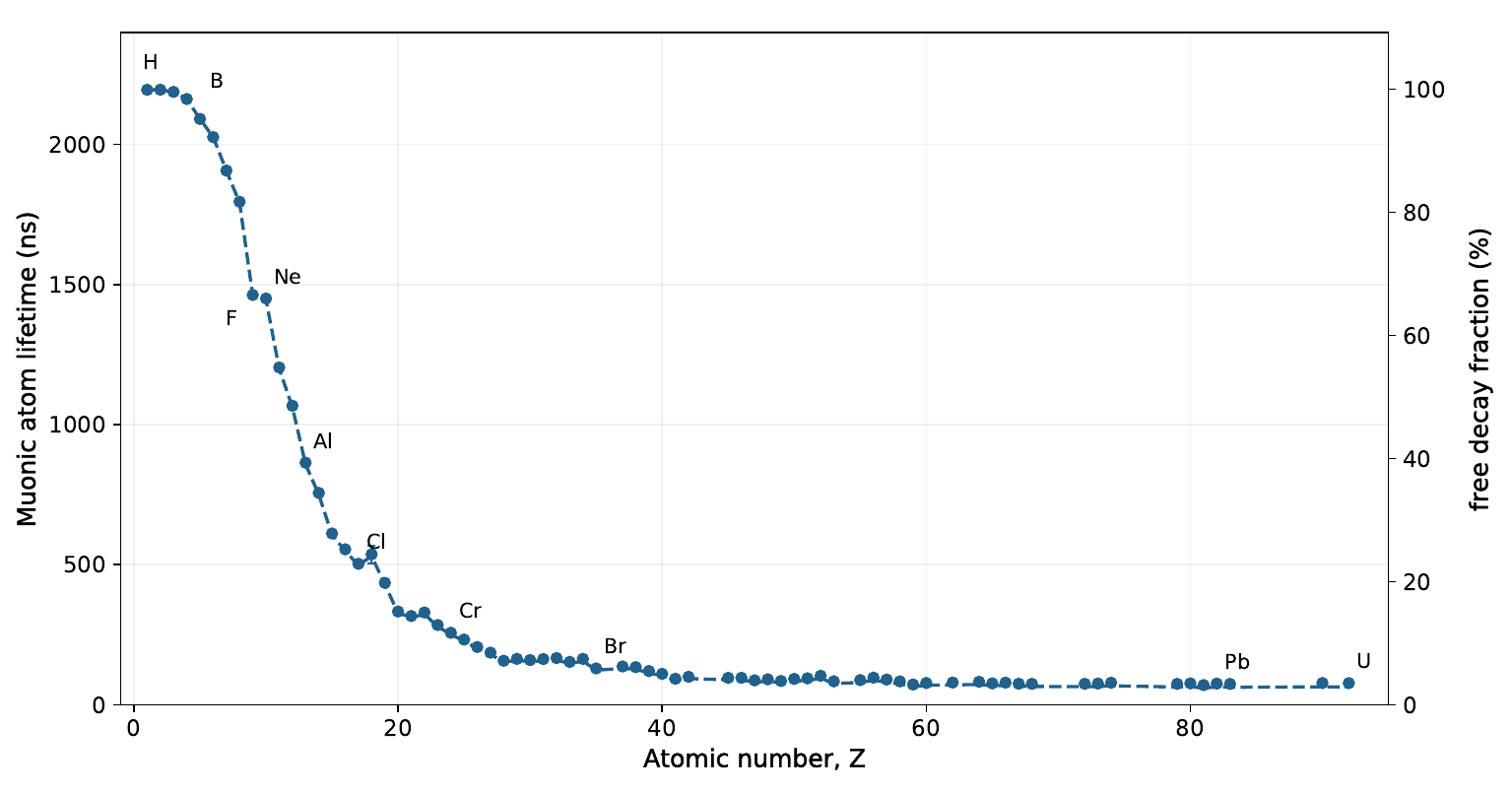}
\vspace{3mm}
\caption{Lifetime of a muonic atom as a function of the atomic number $Z$ of the target nucleus. The right-hand axis shows the free-decay fraction, including the correction due to the Huff factor $Q$. The plot was generated using data from \citet{PhysRevC.35.2212}. Where data for multiple isotopes are available, the lifetime is averaged using their natural abundances as weights.}
\label{fig:muonicatomlifetime}
\end{figure}

Measurements of muonic-atom lifetimes for various nuclei are summarized in \citet{PhysRevC.35.2212}. The dependence of the muonic-atom lifetime on atomic number is shown in \Cref{fig:muonicatomlifetime}.

The muonic-atom lifetime is also important for \mue\ conversion searches. As discussed later, next-generation \mue\ conversion experiments are designed to use pulsed muon beams and search for candidate events in a delayed time window after each pulse in order to suppress beam-related backgrounds. Because high-$Z$ targets have shorter muonic-atom lifetimes, the choice of target material influences the signal acceptance in this delayed time window.

\subsection{Summary of Phenomenological Features}

The quantities introduced in this section define the basic phenomenological features of \mue\ conversion: a monoenergetic electron
from a muonic atom, a target-dependent signal-electron energy, and a time distribution governed by the muonic-atom lifetime. These features provide the basis for the experimental search strategies discussed in later sections.

%
%
%
\section{Theoretical Framework for Experimental Interpretation} 
\label{sec:theory}

The theoretical study of \mue\ conversion has evolved over several decades. Early works included rate estimates based on photonic interactions~\citep{Weinberg:1959zz}, a study of both photonic and nonphotonic interactions~\citep{Marciano:1977cj}, and a quantitative analysis of coherent \mue\ conversion and its dependence on atomic number $Z$~\citep{Shanker:1979ap}. Subsequent studies have included nuclear-matrix-element calculations for coherent elastic and inelastic \mue\ conversion~\citep{Kosmas:1990tc,Kosmas:1993ch}, an investigation incorporating relativistic and finite-nuclear-size effects~\citep{Czarnecki:1998iz}, and systematic predictions for a broad range of nuclear targets and CLFV operator structures~\citep{Kitano:2002mt,Cirigliano:2009bz}.

\subsection{Effective Field Theory Description}
\label{sec:eft}

At energies below the scale of the underlying CLFV physics, \mue\ conversion can be described within an effective field theory (EFT).

At the nucleon level, \mue\ conversion is described by effective interactions involving electrons, muons, photons, and nucleons.\footnote{Interactions with quarks and gluons are matched onto effective interactions with protons and neutrons using nucleon form factors~\citep{Kitano:2002mt,Cirigliano:2009bz}.} The leading effective Lagrangian, $\mathcal{L}_{\mu N \to e N}$, can be written in terms of the dipole operator and four-fermion contact operators involving two charged leptons and two nucleons~\citep{Cirigliano:2017azj,Davidson:2017nrp,Davidson:2018kud}.\footnote{Derivative nucleon operators, which appear in the complete effective-operator description, can also contribute at finite momentum transfer, particularly to spin-dependent \mue\ conversion~\citep{Cirigliano:2017azj,Davidson:2018kud}, but they are not displayed explicitly here.} Following the notation of \citet{Davidson:2018kud}, it is given by
\begin{align}
\mathcal{L}_{\mu N \to e N}
=
-\frac{4G_{\mathrm{F}}}{\sqrt{2}}
\Bigg\{
&\,m_\mu
\big(
C_{D,L}\,\bar e_R \sigma^{\alpha\beta} \mu_L F_{\alpha\beta}
+
C_{D,R}\,\bar e_L \sigma^{\alpha\beta} \mu_R F_{\alpha\beta}
\big)
\nonumber\\
&+
\sum_{N=p,n}
\Big[
\big(
C_{S,L}^{(NN)}\,\bar e P_L \mu
+
C_{S,R}^{(NN)}\,\bar e P_R \mu
\big)\,\bar N N
\nonumber\\
&\qquad+
\big(
C_{P,L}^{(NN)}\,\bar e P_L \mu
+
C_{P,R}^{(NN)}\,\bar e P_R \mu
\big)\,\bar N \gamma_5 N
\nonumber\\
&\qquad+
\big(
C_{V,L}^{(NN)}\,\bar e \gamma^\alpha P_L \mu
+
C_{V,R}^{(NN)}\,\bar e \gamma^\alpha P_R \mu
\big)\,\bar N \gamma_\alpha N
\nonumber\\
&\qquad+
\big(
C_{A,L}^{(NN)}\,\bar e \gamma^\alpha P_L \mu
+
C_{A,R}^{(NN)}\,\bar e \gamma^\alpha P_R \mu
\big)\,\bar N \gamma_\alpha \gamma_5 N
\nonumber\\
&\qquad+
\big(
C_{T,L}^{(NN)}\,\bar e \sigma^{\alpha\beta} P_L \mu
+
C_{T,R}^{(NN)}\,\bar e \sigma^{\alpha\beta} P_R \mu
\big)\,\bar N \sigma_{\alpha\beta} N
\Big]
\Bigg\}
+ \mathrm{h.c.},
\end{align}
where the coefficients $C$ are Wilson coefficients that parameterize the strengths of the corresponding effective interactions. Their subscripts specify the operator structure: $D$, $S$, $P$, $V$, $A$, and $T$ denote dipole, scalar, pseudoscalar, vector, axial-vector, and tensor interactions, respectively. The labels $L$ and $R$ indicate the chirality selected by the charged-lepton bilinear, while the superscript $(NN)$ specifies the nucleon bilinear, with $N=p,n$. The quantities $P_{L,R}$ are the left- and right-handed chiral projectors, and $G_{\mathrm{F}}$ is the Fermi constant.

The effective operators can be classified according to whether they contribute predominantly to spin-independent (SI) or spin-dependent (SD) nuclear responses. The dipole, scalar, and vector operators mainly induce SI \mue\ conversion, while the pseudoscalar, axial-vector, and tensor operators contribute primarily to SD \mue\ conversion.\footnote{This classification is approximate. Axial-vector and tensor interactions can also contribute to coherent SI \mue\ conversion through finite-momentum-transfer effects and loop-induced mixing~\citep{Cirigliano:2017azj}.}

The dipole operator mediates \mue\ conversion through virtual-photon exchange with the nucleus. Since the photon couples to electric charge, this contribution primarily probes the nuclear charge distribution and is coherently enhanced by a factor of order $Z^2$. For high-$Z$ targets, finite-nuclear-size effects modify this simple scaling.

The scalar and vector four-fermion contact operators can also generate SI \mue\ conversion. 
The corresponding amplitudes are obtained by coherently summing over the nucleons in the target nucleus; when the proton and neutron contributions add constructively, this leads to a rate enhancement of order $A^2$. For high-$Z$ targets, this simple scaling is likewise modified by finite-nuclear-size effects, as discussed above.

By contrast, SD interactions couple primarily to nuclear spin degrees of freedom~\citep{Cirigliano:2017azj,Davidson:2017nrp,Hoferichter:2022mna}. Their rates depend on nuclear spin matrix elements, which are often governed by unpaired nucleons. Therefore, they do not receive the coherent $A^2$ enhancement found in SI \mue\ conversion. Pseudoscalar contributions depend on momentum transfer and can be significantly affected by pion exchange~\citep{Davidson:2017nrp}. The underlying quark- and gluon-level CLFV interactions contributing to SD \mue\ conversion, including pseudoscalar, axial-vector, and gluonic operators, can also mediate CLFV decays of light pseudoscalar mesons, $P\to\mu e$, where $P=\pi^0,\eta,\eta^{\prime}$. Present upper limits from SD \mue\ conversion can imply stronger indirect constraints on these decays than those obtained from direct searches~\citep{Hoferichter:2022mna}. More generally, although SD \mue\ conversion does not benefit from the coherent enhancement of SI \mue\ conversion, it provides complementary sensitivity to different CLFV interactions.

Finally, the classification of SI and SD components in \mue\ conversion is analogous to that used in weakly interacting massive particle (WIMP) direct detection, since both processes can involve SI and SD interactions with nuclei~\citep{Engel:1992bf,Jungman:1995df,Fitzpatrick:2012ix}.

\subsection{\texorpdfstring{\mue}{mu-e}\ Conversion Ratio and Nuclear Physics Input}
\label{subsec:si-rate-nuclear-input}

The coherent SI contribution to the \mue\ conversion ratio has been studied extensively~\citep{Kitano:2002mt,Cirigliano:2009bz,Davidson:2018kud}.
Following the notation of \citet{Davidson:2018kud}, it can be written as
\begin{equation}
\mathrm{CR}_{\mathrm{SI}}(\mu^- N \to e^- N)
=
\frac{32 G_{\mathrm{F}}^2 m_\mu^5}{\Gamma_{\mathrm{cap}}}
\left[
\left|
C^{(pp)}_{V,R} V^{(p)}
+
C^{(pp)\prime}_{S,L} S^{(p)}
+
C^{(nn)}_{V,R} V^{(n)}
+
C^{(nn)\prime}_{S,L} S^{(n)}
+
C_{D,L} \frac{D}{4}
\right|^2
+
\{L \leftrightarrow R\}
\right] ,
\label{eq:SI-rate}
\end{equation}
where $\Gamma_{\mathrm{cap}}$ is the total muon-capture rate. The coefficients $C^{(pp)}_{V,L/R}$ and $C^{(nn)}_{V,L/R}$ are the effective vector Wilson coefficients for proton and neutron currents, respectively, while $C^{(pp)\prime}_{S,L/R}$ and $C^{(nn)\prime}_{S,L/R}$ are the corresponding effective scalar coefficients.\footnote{The primes on the scalar coefficients indicate that they include the SI contribution induced by tensor operators at finite momentum transfer.} The coefficients $C_{D,L/R}$ are the effective dipole Wilson coefficients. The quantities $D$, $S^{(p,n)}$, and $V^{(p,n)}$ are target-dependent overlap integrals, defined using the notation of \citet{Kitano:2002mt}. The factor $1/4$ multiplying $D$ follows from the dipole-operator normalization used by \citet{Davidson:2018kud}. The notation $\{L\leftrightarrow R\}$ denotes the term obtained by interchanging $L$ and $R$ in the preceding amplitude.

The nuclear- and atomic-physics input to the SI \mue\ conversion rate enters through the overlap integrals $D$, $S^{(p,n)}$, and $V^{(p,n)}$. These quantities are obtained by integrating products of the bound-muon and outgoing-electron wave functions with the relevant nuclear distributions, including finite-nuclear-size and Coulomb effects. The dipole overlap integral $D$ primarily probes the nuclear charge distribution, while $S^{(p,n)}$ and $V^{(p,n)}$ describe the scalar and vector contributions from proton and neutron distributions.

\subsection{Nuclear-Target Dependence of the \texorpdfstring{\mue}{mu-e}\ Conversion Ratio}
\label{subsec:nuclear-target-dependence}

\begin{figure}[tb!]
\centering
	 \includegraphics[width=0.5\textwidth]{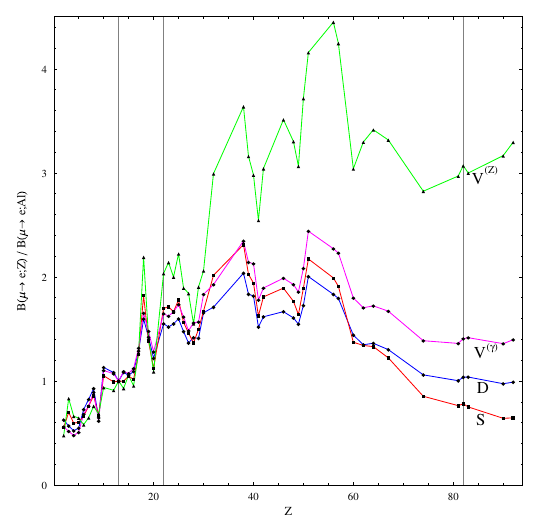}
\caption{
Target dependence of the coherent \mue\ conversion ratio under the assumption that a single SI interaction dominates. The \mue\ conversion ratio is normalized to that for aluminum ($Z=13$) and is shown as a function of the atomic number $Z$. Results are displayed for four benchmark interaction types: dipole ($D$, blue), scalar ($S$, red), photon-mediated vector interaction ($V^{(\gamma)}$, magenta), and $Z$-boson-mediated vector interaction ($V^{(Z)}$, green). Vertical lines indicate $Z=13$ (Al), $Z=22$ (Ti), and $Z=82$ (Pb). Reproduced from \citet{Cirigliano:2009bz}.
}
\label{fig:muec-z-dependence}
\end{figure}

Simple scaling arguments provide useful qualitative intuition for the nuclear-target dependence. For low-$Z$ targets, such arguments indicate that the dipole-dominated \mue\ conversion ratio is approximately proportional to $Z$, although this behavior is substantially modified for high-$Z$ targets.\footnote{In the point-Coulomb picture, the bound-muon probability density in the nuclear region scales approximately as $Z^3$. Together with coherent nuclear enhancement, this motivates an approximate $Z^5$ scaling of the dipole-dominated conversion rate. Since the ordinary muon-capture rate scales approximately as $Z^4$, the conversion ratio then scales approximately as $Z$.} Quantitative comparisons among target nuclei, however, require detailed calculations. Specifically, the target dependence of the SI \mue\ conversion rate is controlled by the overlap integrals appearing in \Cref{eq:SI-rate}. As discussed above, these overlap integrals depend on the nuclear charge distribution, the proton and neutron distributions, and the bound-muon and outgoing-electron wave functions, all of which vary from one target nucleus to another.

The nuclear charge distribution is well constrained by elastic electron scattering and provides the main empirical input for determining the point-proton density distribution, whereas the neutron density distribution is less directly known. \citet{Kitano:2002mt} calculated the coherent SI overlap integrals for 55 nuclei using nuclear charge distributions extracted from elastic electron-scattering data. In their calculations, the neutron density distribution was assumed to be the same as the proton density distribution, although alternative treatments of the neutron distribution were also studied. 

These nuclear-structure inputs determine the overlap integrals $D$, $S^{(p,n)}$, and $V^{(p,n)}$, which enter the \mue\ conversion ratio as target-dependent weights for the contributions of different CLFV operators. Therefore, measurements using different target nuclei can help discriminate among CLFV operators and theoretical models~\citep{Kitano:2002mt,Cirigliano:2009bz}. \Cref{fig:muec-z-dependence} illustrates the target dependence for several representative SI operator structures~\citep{Cirigliano:2009bz}.  The curves correspond to idealized single-operator scenarios and may therefore be regarded as phenomenological benchmarks.

The dipole contribution, denoted by $D$, corresponds to the same photonic dipole operator that also induces \muegamma\ decay. The photon-mediated vector contribution, $V^{(\gamma)}$, can arise from a nondipole photonic operator, whereas $V^{(Z)}$ denotes a weak-neutral-current-like vector contact interaction and can be more sensitive to the neutron component. The scalar contribution, denoted by $S$, represents a scalar contact operator.\footnote{For scalar-dominated scenarios, next-to-leading-order chiral-EFT effects can modify the \mue\ conversion amplitudes at the $\mathcal{O}(10\,\%)$ level~\citep{Cirigliano:2022ekw}.} 

The different curves show that the choice of nuclear target affects both the magnitude of the \mue\ conversion ratio and the relative sensitivity to different CLFV operators. Measurements or limits from multiple target nuclei can therefore provide complementary information on the underlying CLFV operator structure.

Accurate predictions of the target dependence require a proper treatment of nuclear-structure effects in the overlap integrals. The target dependence of coherent \mue\ conversion ratios for 236 stable isotopes with natural abundances above 1\,\% has been studied using different nuclear-density models, with estimates of the resulting nuclear-structure uncertainties~\citep{Heeck:2022wer}. Subsequent work incorporated permanent quadrupole deformation into the relevant nuclear matrix elements using the Barrett-moment method, which allows the nuclear charge distribution, including information on nuclear size and deformation, to be constrained by muonic X-ray data~\citep{Borrel:2024ylg}. This study evaluated coherent \mue\ conversion rates for 236 isotopes and also included estimates of neutron-related matrix elements for even--even nuclei, taking into account quadrupole deformation of the neutron distribution. The calculated \mue\ conversion ratios differ from previous results, particularly for nuclei in regions of large permanent quadrupole deformation.

In parallel, progress has been made in quantifying nuclear uncertainties relevant to coherent SI \mue\ conversion. In particular, residual uncertainties in the nuclear charge distribution have been quantified by propagating elastic electron-scattering data into charge-density parameterizations and the corresponding dipole overlap integrals~\citep{Noel:2024led}. Building on this charge-distribution input, \textit{ab initio} calculations have been used to exploit correlations among charge, point-proton, and point-neutron radii and densities~\citep{Heinz:2024cwg}. This approach yields the scalar and vector overlap integrals $S^{(p,n)}$ and $V^{(p,n)}$ together with full covariance matrices for $^{27}\mathrm{Al}$, $^{48}\mathrm{Ca}$, and $^{48}\mathrm{Ti}$. These results provide nuclear inputs with well-characterized uncertainties for interpreting SI \mue\ conversion searches. The overlap integrals obtained by \citet{Heinz:2024cwg} are reasonably consistent with the earlier results of \citet{Kitano:2002mt}, while providing explicit uncertainty estimates and covariance matrices.

For theoretical-model discrimination, it is useful to consider which nuclear target should be used after an initial search or measurement with a given nucleus. As shown in \Cref{eq:SI-rate}, \mue\ conversion in different target nuclei is sensitive to different combinations of Wilson coefficients, weighted by target-dependent quantities such as $D$, $S^{(p,n)}$, and $V^{(p,n)}$. For a fixed operator basis, these target-dependent weights define vectors in the space of Wilson coefficients. Combining measurements using nuclei with sufficiently different target vectors can therefore improve sensitivity to the operator structure of the CLFV interaction~\citep{Davidson:2018kud}.

For example, if an initial measurement is performed with aluminum, as planned for next-generation experiments, a subsequent measurement with a substantially different target, potentially including a higher-$Z$ nucleus, could provide valuable complementary information on the underlying SI interaction. 

Beyond the dipole, scalar, and vector interactions appearing in \Cref{eq:SI-rate}, additional operators can generate distinctive target dependences. For example, CLFV diphoton contact operators can induce coherent \mue\ conversion through the electric field of the nucleus. In particular, for CLFV diphoton operators involving $F_{\mu\nu}F^{\mu\nu}$, current limits on \mue\ conversion can provide stronger constraints than direct searches for $\mu\to e\gamma\gamma$~\citep{Davidson:2020ord}. Since these contributions involve overlaps with the squared nuclear electric field rather than only the standard SI overlap integrals, they provide another example of how measurements with different nuclear targets can probe nonstandard CLFV interactions.

A more systematic treatment of target dependence in elastic \mue\ conversion is provided by nonrelativistic effective theory (NRET)~\citep{Rule:2021oxe,Haxton:2022piv}. This framework describes low-energy CLFV interactions using a basis of NRET operators defined at the nucleon level. Nuclear-structure effects are incorporated by evaluating the corresponding operators between nuclear states and organizing the resulting matrix elements through a multipole expansion. The elastic \mue\ conversion rate can then be expressed in terms of combinations of coefficients of the NRET operators and six target-dependent nuclear response functions, with some contributions receiving coherent enhancement. These nuclear response functions can be evaluated using nuclear-structure inputs such as advanced shell-model calculations. The NRET coefficients can, in turn, be matched to higher-scale EFT descriptions through an EFT tower~\citep{Haxton:2024lyc}. NRET is therefore complementary to the standard approach based on scalar, vector, and dipole overlap integrals and provides another systematic way to relate \mue\ conversion rates across different targets and to identify the combinations of NRET coefficients that can be constrained.

\subsection{Inelastic \texorpdfstring{\mue}{mu-e}\ Conversion}

Inelastic \mue\ conversion, in which the final nucleus is left in an excited state, can provide information complementary to elastic \mue\ conversion to the nuclear ground state. Transitions to low-lying excited states produce additional peaks in the \mue\ conversion-electron spectrum below the elastic ground-state peak. The energy separations are determined by the corresponding nuclear excitation energies, as indicated in \Cref{eq:inelastic}.

Earlier nuclear-structure studies investigated both inelastic and coherent elastic \mue\ conversion~\citep{Chiang:1993xz,Kosmas:1997eca,Kosmas:2001ia}. Subsequent target-specific calculations evaluated these contributions for experimentally relevant nuclei such as $^{27}\mathrm{Al}$ and $^{48}\mathrm{Ti}$~\citep{Siiskonen:2000nz}, while later work analyzed the multipole decomposition of the \mue\ conversion strength in heavy nuclei such as $^{208}\mathrm{Pb}$~\citep{Civitarese:2019cds}. More recently, the NRET framework discussed above has been extended to inelastic \mue\ conversion~\citep{Haxton:2024ecp,Haxton:2024amf}. The inelastic extension introduces additional nuclear response functions and can probe CLFV operators that are suppressed or inaccessible in elastic \mue\ conversion because of nuclear selection rules.

In experiments, the inelastic peaks may be broadened or only partially resolved because of the finite detector momentum resolution. Nevertheless, inelastic \mue\ conversion can provide complementary information on the underlying CLFV operator structure. In particular, $^{27}\mathrm{Al}$ has been identified as a promising target for obtaining such information~\citep{Haxton:2024ecp,Haxton:2024amf}.

\subsection{Relation to Other Muon CLFV Observables}

As discussed in \Cref{sec:threeCLFVprocesses}, the representative muon CLFV processes (\muegamma\ decay, \mueee\ decay, and \mue\ conversion) probe different combinations of effective CLFV interactions. The decay \muegamma\ is primarily sensitive to dipole operators, while \mueee\ probes four-lepton contact interactions as well as dipole contributions mediated by an off-shell photon. By contrast, \mue\ conversion is sensitive not only to dipole interactions but also to contact interactions involving quarks or, at the nuclear level, nucleons. Scalar, pseudoscalar, vector, axial-vector, tensor, and dipole operators can contribute through different nuclear responses and with different target dependences. Comparisons among these three processes therefore provide complementary information on the underlying CLFV operator structure.

From an EFT viewpoint, these processes are also correlated through the scale dependence of their Wilson coefficients. If new CLFV physics is generated at a high-energy scale, the corresponding effective interactions must be evolved down to the low-energy scale relevant for muon experiments~\citep{Crivellin:2017rmk,Davidson:2016edt}. During this renormalization-group evolution, quantum corrections can mix different operators. Consequently, an operator generated at the high-energy scale can induce additional operator structures at low energies. For example, lepton--quark four-fermion contact interactions can contribute indirectly to \muegamma\ decay through their mixing into dipole operators. These correlations induced by renormalization-group evolution can be accounted for when comparing constraints from \muegamma\ decay, \mueee\ decay, and \mue\ conversion~\citep{Davidson:2020hkf,Davidson:2022nnl}.

From a model-specific viewpoint, particular scenarios can predict characteristic correlations among \muegamma\ decay, \mueee\ decay, and \mue\ conversion. Seesaw models provide one example. In low-scale Type-I seesaw scenarios with heavy-neutrino mixing, these processes are induced at the loop level, and their relative rates depend on the heavy-neutrino masses and mixings. For a quasi-degenerate spectrum, the ratios among the three muon CLFV processes can probe the characteristic heavy-neutrino mass scale, while next-generation \mue\ conversion searches can provide strong sensitivity over a broad mass range~\citep{Alonso:2012ji}. In Type-III seesaw models, tree-level CLFV interactions can enhance \mue\ conversion and \mueee\ decay relative to the loop-induced \muegamma\ decay~\citep{Abada:2007ux}. Such complementarity may therefore help distinguish the underlying origin of CLFV~\citep{Hambye:2013jsa}.

%
%
%
\section{Experimental Basics for \texorpdfstring{\mue}{mu-e}\ Conversion Searches}
\label{sec:basics}

\subsection{Experimental Advantages and Challenges of \texorpdfstring{\mue}{mu-e}\ Conversion}
\label{sec:advantages-challenges}

The \mue\ conversion process has several distinctive experimental features compared with other representative muon CLFV processes, such as \muegamma\ and \mueee\ decays. These features motivate the basic design principles of \mue\ conversion searches.

First, the signal electron in \mue\ conversion has an energy of approximately $105\,\mathrm{MeV}$. This lies well above the dominant part of the DIO-electron spectrum in a muonic atom. Most DIO electrons can therefore be separated efficiently from the signal region. This helps reduce detector occupancy and related experimental challenges while maintaining high signal acceptance. The small high-energy tail near the endpoint energy of the DIO spectrum, however, remains an intrinsic background, as discussed later.

This situation differs from searches for \muegamma\ and \mueee\ decays. In \muegamma\ searches, the signal positron energy lies at the endpoint of ordinary free-muon decay, $E_{e^+}^{\max}\simeq m_\mu/2\simeq 52.8\,\mathrm{MeV}$. In \mueee\ searches, the final-state leptons span a broad energy range overlapping the ordinary muon-decay spectrum.  In both cases, particles from ordinary muon decays can contribute to backgrounds and detector occupancy.

Second, the \mue\ conversion process produces a single monoenergetic electron and therefore does not require coincidence measurements among multiple final-state particles. As a result, it is not subject to the limitations imposed by accidental-coincidence backgrounds, which are important in \muegamma\ and \mueee\ decay searches.\footnote{Accidental-coincidence backgrounds arise when unrelated particles from independent muon decays or other processes occur within the same time window and mimic a signal event.} Such accidental-coincidence backgrounds typically increase rapidly with the instantaneous muon intensity. For example, two-fold accidental coincidences from independent muon-induced events scale approximately as $R^2$, where $R$ denotes the instantaneous muon beam intensity. The absence of accidental-coincidence backgrounds allows \mue\ conversion searches to make effective use of very intense muon beams; in next-generation experiments, the stopped-muon rates under consideration can reach the order of $10^{11}$~muons/s, depending on the experimental concept.

At the same time, the simplicity of the signal also imposes a limitation. Since the experimental signature consists only of a single monoenergetic electron, there are fewer kinematic constraints than in multi-particle final states. Several processes can therefore mimic the signal if they produce an electron reconstructed in the signal region. For instance, high-intensity muon beams can introduce beam-related backgrounds, while muons stopped in the target and surrounding materials can generate additional particles through muon capture and subsequent secondary interactions. Background suppression is therefore a central requirement for \mue\ conversion searches. The relevant background sources and their suppression strategies are discussed in detail in \Cref{sec:backgrounds}.

These advantages and limitations determine the basic experimental strategy of \mue\ conversion searches. 

\subsection{Experimental Principles of \texorpdfstring{\mue}{mu-e}\ Conversion Searches}
\label{sec:principle}

The basic principle of \mue\ conversion searches is to stop a large number of negative muons in a target and search for the monoenergetic electron produced in \mue\ conversion.

The search sensitivity depends critically on the number of useful stopped negative muons. High-intensity proton beams are therefore required to produce a large number of secondary pions at the pion-production target, and the decays of these pions provide tertiary muon beams. For production of low-momentum pions, the pion yield per unit proton-beam power is roughly independent of the proton energy over a limited multi-GeV range~\citep{Ankenbrandt:1999cta,PhysRevSTAB.13.111001}. Under this approximation, the pion-production rate scales with the proton-beam power, which is given by the product of proton energy and proton current. Proton-beam power is therefore a useful first-order measure of the potential pion-production rate, although the actual yield also depends on the target configuration and beamline acceptance.

The proton energy is also important for efficient negative-muon production. At low proton energies, negative-pion production in proton--nucleus interactions is generally suppressed relative to positive-pion production~\citep{Cochran:1972pq}. The negative-pion yield increases with proton energy in the multi-GeV region, and proton energies of several GeV are therefore favorable for efficient negative-pion and negative-muon production~\citep{HARP:2008jan,PhysRevSTAB.13.111001}.

The number of muons that stop in the muon-stopping target and contribute to the accepted search sample should be large. However, this requirement must be balanced against the need to preserve the monoenergetic character of the signal. A thicker stopping target increases the number of stopped muons, but it also increases the energy loss and multiple scattering of \mue\ conversion electrons as they emerge from the target. These effects broaden the reconstructed momentum distribution and produce a tail toward lower energies, reducing the signal acceptance. Therefore, low-energy muons that can be stopped efficiently in a thin target should be used, while the amount of target material traversed by outgoing electrons should be minimized.

The signal region is defined primarily by the reconstructed electron momentum. The detector must therefore provide excellent momentum resolution for the signal electron. A low-mass tracking detector is essential because multiple scattering of the signal electron in detector material can degrade the momentum measurement. 

The very large sample of stopped muons also produces secondary particles at high rates through ordinary muon decay and muon capture. The detector geometry and magnetic-field configuration are therefore designed to reduce the rate at which low-momentum particles, such as DIO electrons, enter the sensitive tracking volume while preserving high acceptance near the \mue\ conversion signal energy. This helps control detector occupancy under high-rate running conditions.

These basic principles apply to both past and future \mue\ conversion searches. Their implementation depends on the available experimental conditions. 

\subsection{Backgrounds and Suppression Strategies}
\label{sec:backgrounds}

Background suppression and rejection are critical elements of \mue\ conversion searches. Since the signal consists of a single electron, a variety of background processes can mimic the signal if they produce an electron reconstructed in the signal region. The relative importance of these backgrounds depends on the intended sensitivity.

The principal backgrounds can be classified into three broad categories: intrinsic physics backgrounds associated with stopped muons, beam-related backgrounds produced by particles accompanying the muon beam, and backgrounds originating from cosmic rays or other external sources. The major sources in each category are discussed below in detail.

\subsubsection{Muon-Induced Intrinsic Backgrounds}

Intrinsic physics backgrounds arise from negative muons stopped in the muon-stopping target. The two primary sources of signal-like electrons are muon decay in orbit and radiative muon capture. These are discussed first. Other particles produced following ordinary muon capture in nuclei can also contribute to detector occupancy and radiation damage, as discussed at the end of this subsection.

\paragraph{Muon Decay in Orbit}
\label{sec:dio}

\begin{figure}[b!]
\centering
	 \includegraphics[width=\textwidth]{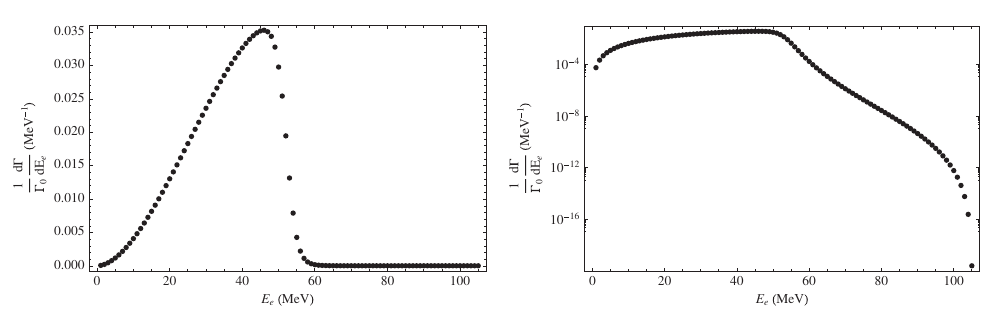}
  \caption{DIO-electron spectrum for aluminum. The left plot shows the spectrum on a linear scale, while the right plot shows it on a logarithmic scale. Reproduced from \citet{Czarnecki:2011mx}.}
  \label{fig:diospectrum-aluminium}
\end{figure}

A primary intrinsic physics background is muon decay in orbit (DIO). This process occurs when a negative muon bound in the $1\mathrm{s}$ state of a muonic atom undergoes ordinary muon decay. As discussed in \Cref{sec:timedistribution}, bound-state effects modify the energy spectrum of the DIO electron. The DIO-electron spectrum has been studied extensively, from early calculations~\citep{Uberall:1960zz,Haenggi:1974hp,Herzog:1980my,Watanabe:1993emp} to detailed endpoint-spectrum calculations developed for \mue\ conversion searches~\citep{Czarnecki:2011mx}.

Because the muon is bound to the nucleus and the nucleus can absorb recoil momentum, the DIO-electron spectrum extends up to the signal energy of \mue\ conversion. Near this endpoint, the two neutrinos carry only a small amount of energy, while the electron and recoiling nucleus are emitted approximately back-to-back. This kinematic configuration resembles the two-body final state of the \mue\ conversion signal. The endpoint spectrum falls steeply, approximately as
\begin{equation}
\frac{d\Gamma_{\rm DIO}}{dE_e} \propto (E_{\mathrm{end}}-E_e)^5 ,
\end{equation}
where $E_e$ is the DIO-electron energy and $E_{\mathrm{end}}$ is the endpoint energy. The DIO-electron spectrum for aluminum is shown in \Cref{fig:diospectrum-aluminium}.

This steep fall is advantageous because most DIO electrons have much lower energies than the \mue\ conversion signal. However, the high-energy endpoint tail remains an irreducible background. The DIO contribution in the signal region is therefore highly sensitive to the electron momentum resolution and the definition of the signal window.

For experimental searches, the relevant quantity is the integrated probability for a DIO electron to lie above the lower edge of the signal region. This can be written as
\begin{equation}
P_{\rm DIO}(x)
=
\int^{E_{\mathrm{end}}}_{x}
\frac{1}{\Gamma_0}
\frac{d\Gamma_{\rm DIO}}{dE_e}
\,dE_e ,
\end{equation}
where $x$ is the lower edge of the signal energy window, and $\Gamma_0$ is the free-muon decay rate. Since the endpoint spectrum approximately scales as $(E_{\mathrm{end}}-E_e)^5$, the integrated probability scales as $(E_{\mathrm{end}}-x)^6$. This strong dependence illustrates why excellent momentum resolution is essential.

\begin{figure}[tb!]
\centering
	 \includegraphics[width=0.6\textwidth]{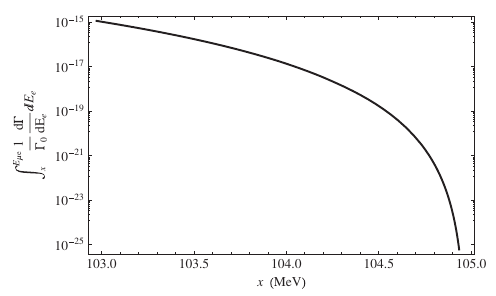}
\caption{Integrated rate of DIO events in aluminum with electron energy larger than $x$, normalized to the free-muon decay rate. Reproduced from \citet{Czarnecki:2011mx}.}
  \label{fig:integrated-dio}
\end{figure}

\Cref{fig:integrated-dio} shows $P_{\rm DIO}(x)$ as a function of the lower energy threshold $x$, normalized to the free-muon decay rate~\citep{Czarnecki:2011mx}. For aluminum, the integrated DIO probability is reduced to the level of $\mathcal{O}(10^{-17})$ when the lower edge of the signal energy region is placed near $103.5\,\mathrm{MeV}$ or higher, illustrating the stringent momentum-window requirement needed to keep the DIO background at an acceptable level. The integrated rates shown in \Cref{fig:integrated-dio} do not include radiative corrections. These corrections, evaluated in a later paper and discussed below, were found to reduce the integrated rates by about 11\,\% and 15\,\% for endpoint energy bins of $1\,\mathrm{MeV}$ and $0.15\,\mathrm{MeV}$, respectively~\citep{Szafron:2015kja}. 

More refined calculations of the DIO endpoint spectrum include radiative corrections, vacuum-polarization effects, and nuclear charge distributions~\citep{Szafron:2016cbv,Szafron:2015kja,Heeck:2021adh}. These effects modify both the normalization and the shape of the endpoint tail. Near the endpoint, the spectrum with the leading radiative corrections can be parameterized as
\begin{equation}
\frac{1}{\Gamma_0}\frac{d\Gamma}{dE_e}\Bigg |_{E_e\sim E^{\prime}_{\rm end}}
\simeq
B \left( E^{\prime}_{\rm end} \right) ^{5}
\left(
1 - \frac{E_e}{E^{\prime}_{\rm end}}
\right)^{5+\delta} ,
\end{equation}
where $B$ determines the normalization of the endpoint tail, $E^{\prime}_{\rm end}$ is the endpoint energy including vacuum-polarization effects, and $\delta\simeq0.023$ accounts for soft-photon radiation.\footnote{The soft-photon exponent is
$\delta = \frac{\alpha}{\pi}\left(2\log\frac{2m_\mu}{m_e}-2\right)\simeq 0.023$~\citep{Szafron:2015kja}.}
The same vacuum-polarization shift applies to the DIO endpoint and to the \mue\ conversion signal energy. Thus, the shifted DIO endpoint, $E^{\prime}_{\rm end}=E_{\rm end}+\Delta E_{\rm VP}$, coincides with the corrected \mue\ conversion-signal energy, $E^{\prime}_{\mu e}$, where $\Delta E_{\rm VP}$ is given in \Cref{eq:vacuumpolarization}.

\begin{figure}[tb]
\centering
	 \includegraphics[width=0.6\textwidth]{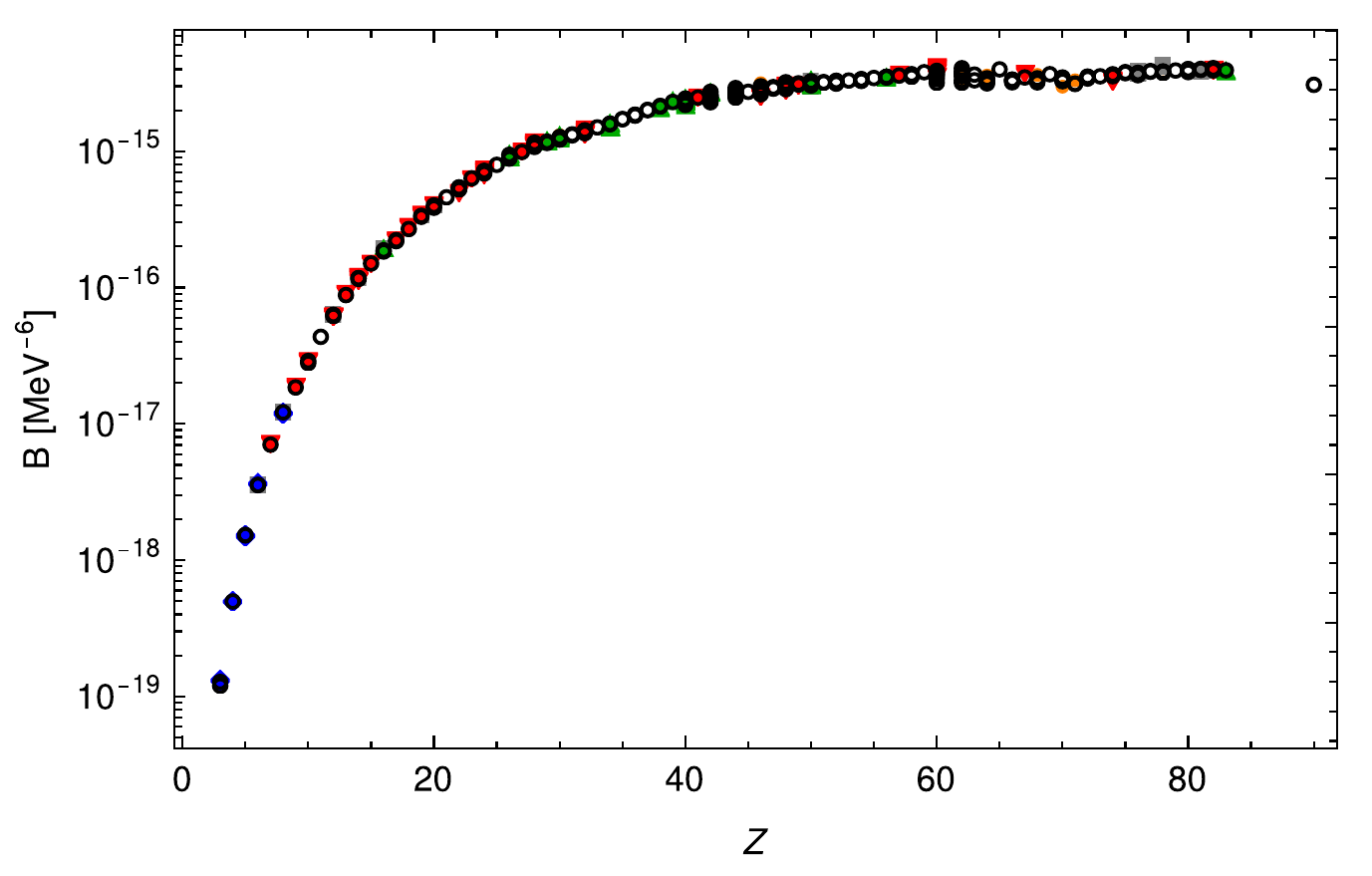}
\caption{Atomic-number dependence of the coefficient $B$ in the DIO-electron endpoint spectrum. Reproduced from \citet{Heeck:2021adh}, licensed under CC BY 4.0.}
  \label{fig:diobcoefficient}
\end{figure}

The coefficient $B$ increases with atomic number $Z$, reflecting stronger bound-state and Coulomb-field effects in high-$Z$ nuclei. This enhances the relative population of DIO electrons near the endpoint. The calculated $Z$ dependence of $B$, obtained using nuclear charge distributions, is shown in \Cref{fig:diobcoefficient}~\citep{Heeck:2021adh}. The dominant uncertainty in $B$ arises from the nuclear charge distribution and is currently about $5$--$10\,\%$.

For high-$Z$ nuclei, two effects compete. The total number of DIO electrons is reduced because ordinary muon capture dominates over DIO. At the same time, the endpoint tail is enhanced by the larger value of $B$. These effects can partially compensate for each other, so the DIO background in the signal region does not necessarily decrease simply by choosing a high-$Z$ target.

Experimentally, the DIO background is controlled primarily by achieving excellent electron momentum resolution. For \mue\ conversion electrons near $105\,\mathrm{MeV}$, the resolution of a tracking detector is often limited by multiple scattering in the stopping target and detector material. Minimal detector material, a thin stopping target, and careful treatment of energy loss are therefore essential, together with precise momentum calibration. Since the DIO rate entering the signal window scales very steeply with the effective width of the endpoint region, even modest degradation of the momentum resolution can significantly increase the expected background.

Special care is also required to avoid contamination from unintended low-$Z$ materials near the stopping target. 
Such materials can produce DIO spectra with higher endpoint energies than those of the nominal stopping-target material, potentially introducing poorly controlled high-energy electron backgrounds.

\paragraph{Radiative Muon Capture}
\label{sec:rmc-background}

Radiative muon capture (RMC) is another intrinsic background associated with stopped negative muons. The RMC process without nucleon emission can be written as
\begin{equation}
\mu^{-}+N(A,Z)\rightarrow \nu_{\mu}+N(A,Z-1)+\gamma,
\label{eq:rmc}
\end{equation}
where the final nucleus $N(A,Z-1)$ may be left either in its ground state or in an excited state.
Additional processes with nucleon emission can also occur. Radiative muon capture has been measured for several nuclear targets~\citep{Armstrong:1992sn,Bergbusch:1999ms}.  The partial branching ratios for photons above
$57\,\mathrm{MeV}$, $R_\gamma$, are typically of order $10^{-5}$ relative to ordinary
nonradiative muon capture. For example, $R_\gamma=(1.40\pm0.11)\times10^{-5}$ for aluminum and $(1.30\pm0.12)\times10^{-5}$ for titanium.

The emitted photon can produce an $e^+e^-$ pair either externally, by converting in the stopping target or nearby detector material, or internally, through virtual-photon conversion. In realistic detector geometries, the amount of material near the stopping target is not negligible, so internal and external conversion contributions can be comparable. Electrons can also be produced through Compton scattering. Such events become backgrounds when the produced electron carries sufficiently high energy to enter the signal region.  

The RMC background can be reduced by choosing an appropriate stopping-target material. For a target nucleus with mass number $A$ and atomic number $Z$, the maximum photon energy in RMC is approximately
\begin{equation}
E_{\rm RMC}^{\rm max}
\approx
m_{\mu}-B_{\mu}-E_{\rm recoil}+M(A,Z)-M(A,Z-1),
\end{equation}
where $m_{\mu}$ is the muon mass, $B_{\mu}$ is the binding energy of the muon in the $1\mathrm{s}$ state of a muonic atom, $E_{\rm recoil}$ is the nuclear recoil energy, and $M(A,Z)$ and $M(A,Z-1)$ are the masses of the initial and final nuclei.
This expression corresponds to the idealized case in which the final nucleus remains in its ground state and no nucleons are emitted. Excitation of the final nucleus or nucleon emission lowers the available photon energy.

Such high-energy photons can produce electrons near the signal region. The RMC background is kinematically suppressed if the endpoint photon energy is sufficiently below the \mue\ conversion signal energy. This favors target nuclei for which $M(A,Z)-M(A,Z-1)<0$, so that the nuclear mass difference lowers the RMC endpoint relative to the \mue\ conversion signal energy. For example, $M(^{27}\mathrm{Al})-M(^{27}\mathrm{Mg}) \simeq -3.1\,\mathrm{MeV}$, and $M(^{48}\mathrm{Ti})-M(^{48}\mathrm{Sc}) \simeq -4.5\,\mathrm{MeV}$. This kinematic separation is one of the reasons why aluminum and titanium have been attractive stopping-target materials for suppressing the RMC background.

\paragraph{Particle Emission after Ordinary Muon Capture}

Ordinary muon capture in nuclei also produces secondary particles such as protons, neutrons, photons, and nuclear fragments. Although these particles do not usually produce signal-like electrons directly, they can increase detector hit rates and degrade detector performance. 

Protons are particularly important because they can deposit large amounts of energy in tracking detectors, potentially saturating the detector response or causing temporary dead time. The proton-emission probability following muon capture is not negligible. For aluminum and titanium, it is about $0.03$ per muon capture for protons with kinetic energies in the range of approximately $4$--$20\,\mathrm{MeV}$~\citep{AlCap:2021msk}. Proton-induced detector effects can be mitigated by charge and momentum selection to prevent protons from reaching the detector, or by placing thin absorbers, such as foils or detector entrance windows, upstream of the tracking volume to reduce the proton flux.

Neutron emission following ordinary muon capture is also important. For $^{27}\mathrm{Al}$, recent activation measurements and previous neutron-multiplicity evaluations indicate that neutron-emission processes make a major contribution to nuclear de-excitation following muon capture, with an average multiplicity of order one neutron per capture~\citep{Mizuno:2025kxh}. Neutrons can penetrate shielding and detector materials more easily than charged particles, contributing to detector hit rates and radiation damage. Their impact depends on the muon-stopping target material, detector geometry, and shielding configuration. For this reason, neutron production and transport must be evaluated using experiment-specific simulations and, where possible, constrained by dedicated neutron measurements.

\subsubsection{Beam-Related Backgrounds}

Beam-related backgrounds arise from particles accompanying the muon beam, particularly residual pions and electrons. They can produce signal-like events either directly or through secondary interactions in the muon-stopping target or beamline. Their suppression requires a combination of beam control, shielding, and timing cuts.

\paragraph{Beam-Pion Background}
\label{sec:beampion}

Beam pions constitute a potentially serious background if they reach the stopping target. Negative pions that stop in the target are captured promptly by nuclei and can undergo radiative pion capture (RPC),
\begin{equation}
\pi^{-}+N\rightarrow X+\gamma ,
\end{equation}
where $X$ denotes the final nuclear state, possibly including emitted nucleons or fragments. RPC photons can produce signal-like electrons through photon conversion into an $e^+e^-$ pair or through Compton scattering, as in the case of RMC. However, RPC backgrounds are kinematically more problematic because the pion mass is larger than the muon mass, so RPC photons can populate an energy range that overlaps the \mue\ conversion signal region. Therefore, unlike RMC, whose contribution can be reduced by choosing an appropriate target nucleus as discussed earlier, RPC backgrounds cannot be suppressed by target choice.

The photon-emission probability in RPC is at the percent level and depends only weakly on the nuclear charge $Z$.  Previous experimental and theoretical studies show probabilities of approximately $2\,\%$ for light and medium nuclei such as C, O, and Ca, with photon spectra extending over a broad range, roughly from several tens of MeV up to about $140\,\mathrm{MeV}$~\citep{Bistirlich:1972jy}. These features make RPC a potentially serious beam-related background if pions arrive at the stopping target during the measurement period.

Pion-induced backgrounds can be suppressed by exploiting the different properties of pions and muons in the beamline. Past experiments used pion range absorbers, taking advantage of the different penetration ranges of pions and muons in material. In next-generation pulsed-beam experiments, pion suppression relies mainly on the short pion lifetime relative to the muon lifetime, together with strong suppression of beam particles arriving outside the intended beam pulses. These methods are discussed in detail in \Cref{sec:timing-extinction}.

\paragraph{Beam Electrons and Decays in Flight}

Beam electrons originate mainly from electromagnetic showers induced by proton interactions in the pion-production target, including photons from $\pi^0$ decays. If transported to the stopping target, these electrons can scatter in the target or nearby materials and mimic a \mue\ conversion signal when their reconstructed momentum falls within the signal region.

Additional beam-related backgrounds arise from decays in flight. Muons and pions that decay in flight can produce energetic electrons. 
For a signal region beginning at around $103\,\mathrm{MeV}/c$, the corresponding parent-momentum thresholds are roughly $75\,\mathrm{MeV}/c$ for muons and $55\,\mathrm{MeV}/c$ for pions. Although the pion decay $\pi^- \to e^- \bar{\nu}_e$ is suppressed by its small branching ratio,
$\mathrm{BR}(\pi^- \to e^- \bar{\nu}_e) \simeq 1.2\times10^{-4}$, it can still contribute to the background in high-intensity muon beams. Kaon decays in flight may also contribute, depending on the proton-beam energy.

These backgrounds can be reduced by limiting the momentum acceptance of the beamline, thereby suppressing particles that could decay into signal-like electrons. A delayed measurement window further reduces prompt and early-arriving beam-related background particles, although residual decay-in-flight backgrounds must be evaluated using the experiment-specific beamline geometry and timing structure.

\paragraph{Antiproton Backgrounds}

Antiprotons can be a beam-related background in \mue\ conversion searches when the primary proton-beam energy is sufficiently high. They are produced in hadronic interactions of the proton beam with the pion-production target or nearby material. Their annihilation products, including pions, photons, and charged hadrons, can generate electrons with momenta near the signal region through secondary interactions. This background is relevant for multi-GeV proton beams but is negligible for facilities operating well below the antiproton-production threshold.

A particular concern is that low-momentum antiprotons can travel slowly and arrive at late times, making them difficult to reject by event timing alone. This background can be reduced by limiting both the production and transport of antiprotons through suitable choices of proton-beam energy and by placing antiproton absorbers in the beamline. For instance, since antiproton production has a threshold in the few-GeV range,\footnote{For a free proton target, the threshold kinetic energy for antiproton production in $p+p \to p+p+p+\bar p$ is about $5.6\,\mathrm{GeV}$. For nuclear targets, Fermi motion, binding effects, and multi-nucleon processes smear this threshold and can allow antiproton production to occur somewhat below the free $p+p$ threshold, although the yield remains strongly suppressed near threshold.} the proton-beam energy can be chosen to be sufficiently low while still providing adequate pion production. As antiproton absorbers, multiple thin foils can be placed in the beamline, allowing muons to pass through while antiprotons are substantially slowed down or stopped. These measures help suppress the antiproton background.

The antiproton yield in the backward direction from an $8\,\mathrm{GeV}$ proton beam, as relevant for next-generation experiments, is not well known and is difficult to predict reliably with present hadron-production simulations. Dedicated measurements or conservative background estimates are therefore important for evaluating antiproton-induced backgrounds.

\subsubsection{Cosmic-Ray Backgrounds}

Cosmic-ray-induced backgrounds arise when cosmic-ray muons, or electrons produced by cosmic-ray interactions, are reconstructed as signal-electron candidates. They are among the most important non-beam background sources in \mue\ conversion searches. These backgrounds can be broadly divided into two classes: cosmic-ray muons that interact with material and produce electrons within the detector acceptance, and cosmic-ray muons that enter the detector and are misidentified as \mue\ conversion signal candidates.

Cosmic-ray backgrounds are particularly important because they are not correlated with the beam structure and therefore cannot be suppressed by event timing. Their total contribution is proportional to the live running time of the experiment rather than to the proton-beam power or stopped-muon rate.

To suppress these backgrounds, the detector system and relevant parts of the beamline are typically surrounded by multilayer cosmic-ray veto detectors, such as plastic scintillator counters. These detectors identify incoming cosmic-ray muons and veto the corresponding events. The cosmic-ray veto system must cover a large fraction of the solid angle around the detector and beamline and operate with very high efficiency, typically exceeding $99.99\,\%$, corresponding to an inefficiency of order $10^{-4}$ or less. This performance must be maintained in the presence of beam-related radiation, especially neutrons, which can cause accidental veto signals and detector damage.

Cosmic-ray-induced backgrounds can also be constrained directly using data collected during beam-off periods. Dedicated cosmic-ray runs with the full detector, trigger, and particle-identification systems are important for validating simulations, studying residual backgrounds, and determining whether additional suppression measures are required before or during the physics run.

In summary, background suppression in \mue\ conversion searches requires a coordinated experimental strategy. Important elements include excellent momentum resolution, suitable stopping-target choice, control of beam-related particles, suppression of cosmic-ray backgrounds with veto detectors, and evaluation of secondary-particle production under high-intensity beam conditions. 
These requirements have shaped the design and sensitivity of past searches, which are reviewed in the next section, and continue to guide the design of next-generation experiments discussed later.

%
%
%
\section{Past Experiments}

The first reported experimental search for \mue\ conversion was carried out in 1952 using cosmic-ray muons~\citep{Lagarrigue1952}. Since this pioneering experiment, accelerator-based searches have been performed using several target materials, including sulfur (S, $Z=16$), titanium (Ti, $Z=22$), copper (Cu, $Z=29$), gold (Au, $Z=79$), and lead (Pb, $Z=82$). The historical limits are summarized in \Cref{table:muec-history}.

\begin{table}[bt!]
\setlength{\heavyrulewidth}{1pt}
\centering
\caption{Historical upper limits on the \mue\ conversion ratio in nuclei. For each measurement, the table shows the year, target nucleus, experimental site, upper limit on the \mue\ conversion ratio at 90\,\% C.L., and reference. The reference marked with $\dagger$ was published only in conference proceedings.}
\label{table:muec-history}
\vspace{5mm}
\begin{tabular}{lllll}
\toprule
Year & Target & Site & $\mathrm{CR}$ upper limit at 90\,\% C.L. & Reference \\
\midrule
1952 & Sn, Sb & cosmic rays & $<1 \times 10^{-1}$ & \cite{Lagarrigue1952} \\
1955 & Cu & Nevis & $<5 \times 10^{-4}$ & \cite{PhysRev.100.1490} \\
1960 & Cu & LBL & $<4 \times 10^{-6}$ & \cite{Sard:1960qja} \\
1961 & Cu & CERN & $<5 \times 10^{-6}$ & \cite{PhysRev.122.687} \\
1962 & Cu & CERN & $<2.2 \times 10^{-7}$ & \cite{Conforto:1962zz} \\
1964 & Cu & Liverpool & $<2.2 \times 10^{-7}$ & \cite{BARTLEY1964258} \\
1972 & Cu & SREL & $<1.6 \times 10^{-8}$ & \cite{PhysRevLett.28.1469} \\
1977 & Cu & SIN & $<4 \times 10^{-10}$ & \cite{PhysRevLett.39.1385} \\
1982 & S & SIN & $<7 \times 10^{-11}$ & \cite{BADERTSCHER1982406} \\
1985 & Ti & TRIUMF & $<1.6 \times 10^{-11}$ & \cite{PhysRevLett.55.465} \\
1988 & Ti & TRIUMF & $<4.6 \times 10^{-12}$ & \cite{Ahmad:1988ur} \\
1988 & Pb & TRIUMF & $<4.9 \times 10^{-10}$ & \cite{Ahmad:1988ur} \\
1993 & Ti & PSI & $<4.3 \times 10^{-12}$ & \cite{DOHMEN1993631} \\
1996 & Pb & PSI & $<4.6 \times 10^{-11}$ & \cite{PhysRevLett.76.200} \\
1998 & Ti & PSI & $<6.1 \times 10^{-13}$ & \cite{Wintz:1998aa}$^{\dagger}$ \\
2006 & Au & PSI & $<7 \times 10^{-13}$ & \cite{Bertl:2006te} \\
\bottomrule
\end{tabular}
\end{table}

\subsection{SINDRUM~II Experiment at PSI}

Among the results reported in refereed journals, the most stringent upper limit on \mue\ conversion was obtained by the SINDRUM~II experiment at the Paul Scherrer Institute (PSI) using a gold target. For gold, the \mue\ conversion-electron momentum is about $95.6\,\mathrm{MeV}/c$, and the muonic-atom lifetime is about $72\,\mathrm{ns}$. The data were collected in 2000, and the final result, published in 2006, was
\begin{equation}
\mathrm{CR}(\mu^- \mathrm{Au} \to e^- \mathrm{Au}) < 7 \times 10^{-13},
\end{equation}
at the 90\,\% confidence level (C.L.)~\citep{Bertl:2006te}.

The experiment was carried out at the $\pi$E5 beamline at PSI, where the $590\,\mathrm{MeV}$ quasi-continuous proton beam from the PSI cyclotron was used to produce an intense negative-muon beam.\footnote{The proton beam had a time structure consisting of bursts about $0.3\,\mathrm{ns}$ wide, separated by $19.75\,\mathrm{ns}$, a spacing shorter than the muonic-atom lifetime ($72\,\mathrm{ns}$).}
The muon beam was operated at central momenta of $52$ and $53\,\mathrm{MeV}/c$, which are sufficiently low to eliminate background from muon decay in flight. Pions in the muon beam were suppressed using an $8\,\mathrm{mm}$-thick CH$_2$ degrader installed in the beamline, exploiting the fact that the pion range is about half the muon range.
The muons were then transported to the target through a 9-m-long superconducting transport solenoid, and almost all of the low-momentum pions that passed the degrader decayed in flight before reaching the target.

The SINDRUM~II spectrometer consisted of a superconducting solenoid housing cylindrical detectors arranged around the target. The gold target, located at the center of the spectrometer, was a 65-cm-long tube with a wall thickness of only $75\,\mathrm{mg/cm}^2$, which kept the energy loss of the emitted electrons small. Two plastic-scintillator hodoscopes and a Cherenkov hodoscope provided timing and triggering, and the momenta of electrons emitted from the target were measured with two cylindrical drift chambers. The number of muons stopped in the target was monitored by measuring muonic X-rays from gold with a Ge(Li) detector placed outside the spectrometer coil. The total accumulated sample was $(4.37 \pm 0.32)\times 10^{13}$ stopped negative muons.

Pion-induced backgrounds, in particular radiative pion capture (RPC) in the degrader and pion decay in flight upstream of the degrader, were rejected using their forward-peaked angular distribution and their timing correlation with the cyclotron RF, which arises because these processes are induced by pions arriving in time with the proton bursts.
Cosmic-ray events were rejected on the basis of additional tracks in the detectors. Positrons were recorded together with the electrons and were used to study background processes in which a photon converts into an $e^+e^-$ pair. The remaining electron momentum spectrum below the signal region was dominated by DIO electrons, whose spectrum falls steeply toward its endpoint. The expected signal is a peak at the conversion momentum with a low-momentum tail caused by energy loss in the target. No events were observed in the signal region. The upper limit was obtained from a likelihood analysis of the momentum spectra~\citep{Bertl:2006te}.

%
%
%
\section{Experimental Strategy for Next-Generation Searches}
\label{sec:strategy}

\subsection{Overview}

Next-generation \mue\ conversion experiments are designed to improve sensitivity by roughly four orders of magnitude. The basic experimental principles of \mue\ conversion searches were introduced in \Cref{sec:basics}. Next-generation experiments apply the same principles but require much larger data samples. Previous searches used stopped-muon samples approaching $\mathcal{O}(10^{14})$ and reached sensitivities to \mue\ conversion at the level of $\mathcal{O}(10^{-12})$. For example, the SINDRUM~II search using a gold target accumulated $(4.37 \pm 0.32)\times 10^{13}$ stopped $\mu^-$ and set an upper limit of $7\times10^{-13}$ on the \mue\ conversion ratio~\citep{Bertl:2006te}. Searches aiming at sensitivities around $10^{-16}$, corresponding to a four-order-of-magnitude improvement, require total samples of $\mathcal{O}(10^{18})$ stopped $\mu^-$, although the precise requirement depends on the experimental conditions. Accumulating $10^{18}$ stopped muons corresponds to an average rate of about $10^{11}$ stopped $\mu^-$ per second for a running time of $10^7\,\mathrm{s}$.

This four-order-of-magnitude improvement requires a substantially more advanced experimental approach. The resulting requirements define the basic strategy of next-generation experiments, as described in the following subsections. Specific implementations, such as COMET at J-PARC and Mu2e at Fermilab, are discussed in \Cref{sec:currentfutureexperiments}.

\subsection{High-Intensity Stopped-Muon Production and Transport}
\label{sec:captureandtransport}

Achieving the large samples of stopped muons needed for these experiments requires the use of solenoid pion capture and curved-solenoid muon transport. In this review, we refer to pion capture using a high-field solenoid as solenoid pion capture and to muon transport through a curved solenoid as curved-solenoid muon transport. A high-intensity proton beam strikes a pion-production target, generating charged pions and other secondary particles. A high-field superconducting solenoid surrounding the production target captures charged pions over a large solid angle, many of which decay in flight into muons. The resulting muon beam is then guided toward the muon stopping target through a downstream curved-solenoid muon-transport system, which provides charge and momentum selection. 

\begin{figure}[tb!]
\centering
	 \includegraphics[width=0.6\textwidth]{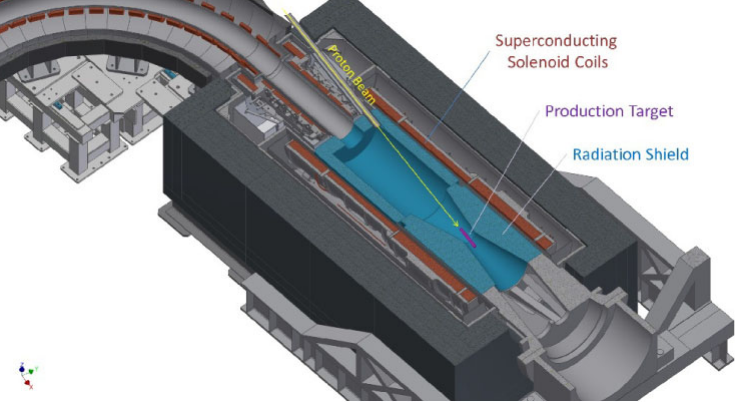}
\caption{Schematic layout of the solenoid pion-capture and muon-transport concept, shown for the COMET configuration. The proton beam enters from above and strikes a cylindrical pion-production target surrounded by a superconducting capture solenoid. Low-energy negative pions emitted backward with respect to the incident proton-beam direction are collected, and negative muons produced from their decays are transported downstream through the solenoidal beamline. Reproduced from \citet{10.1093/ptep/ptz125}, licensed under CC BY 4.0.} 
\label{fig:pion-capture-solenoid}
\end{figure}

As an example of this general pion-capture principle, \Cref{fig:pion-capture-solenoid} shows a schematic configuration of the pion-capture solenoid adopted by COMET. In this design, a long, small-radius cylindrical pion-production target is placed in a solenoidal field. The length of the pion-production target can be of the order of one nuclear interaction length, corresponding to about $40\,\mathrm{cm}$ for graphite and $28\,\mathrm{cm}$ for titanium. The surrounding solenoidal magnets are used to collect as many pions emerging from the long cylindrical pion-production target as possible, in particular those emitted backward with respect to the incident proton-beam direction. This backward selection preferentially captures lower-energy pions, which subsequently decay into low-energy muons suitable for stopping, whereas higher-energy pions are produced predominantly in the forward direction. Further details of the beam optics of solenoid pion capture are given in \Cref{app:pioncapture}.

The curved-solenoid muon-transport system following the solenoid pion-capture system generates a charge- and momentum-dependent transverse displacement of the beam particles.\footnote{In beam optics, momentum-dependent transverse displacement is referred to as dispersion.} The resulting spatial separation between particles of different charges and momenta enables charge and momentum selection using collimators or apertures together with correction dipole fields. As a result, the transport system efficiently delivers an intense $\mu^-$ beam with a momentum distribution suitable for stopping in the muon stopping target, while suppressing oppositely charged particles, other unwanted particles, and beam-related backgrounds. Further details of the beam optics of curved-solenoid muon transport are given in \Cref{app:curvedsolenoid}.

\begin{figure}[tb]
\centering
	 \includegraphics[width=0.6\textwidth]{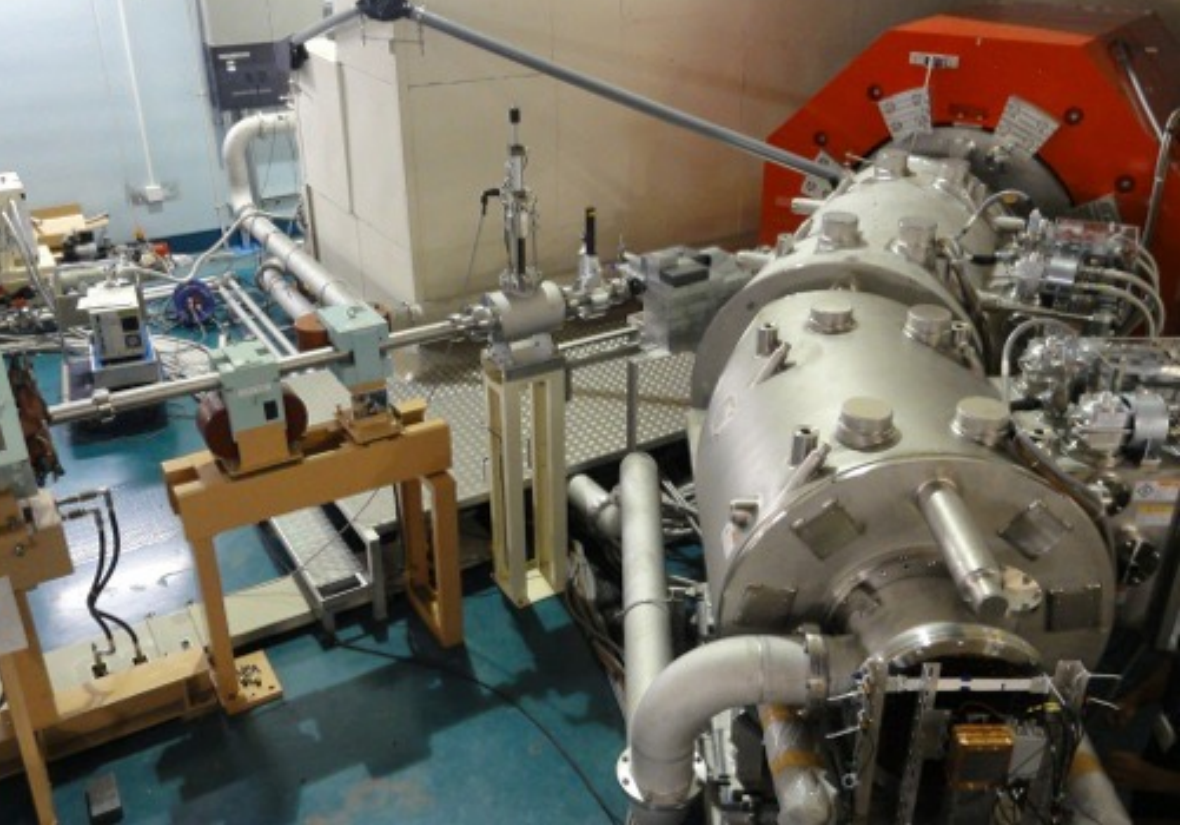}
\caption{Photograph of the MuSIC muon source at the Research Center for Nuclear Physics (RCNP), The University of Osaka. The proton beam from the cyclotron enters from the left and strikes the pion-production target located inside the pion-capture solenoid, which is housed within the red iron yoke visible in the upper right. The resulting muon beam is transported downstream through the curved-solenoid system. Photograph courtesy of RCNP, The University of Osaka.}
\label{fig:music-photo}
\end{figure}

Solenoid pion capture and curved-solenoid muon transport were experimentally studied in 2011 at the MuSIC facility\footnote{MuSIC stands for Muon Science Innovative Channel.} of the Research Center for Nuclear Physics (RCNP), The University of Osaka, formerly known as Osaka University~\citep{Cook:2016sfz}. MuSIC demonstrated the required performance of the solenoid pion-capture and curved-solenoid muon-transport systems, as well as their technical feasibility. A photograph of the MuSIC pion-capture solenoid and muon-transport system is shown in \Cref{fig:music-photo}. Technical details of the MuSIC implementation are summarized in \Cref{app:music}.

\subsection{Timing Strategy and Beam Extinction}
\label{sec:timing-extinction}

A pulsed proton beam is a central feature of next-generation searches for \mue\ conversion. Prompt particles produced by each proton pulse arrive near the beam-pulse time, whereas \mue\ conversion electrons are emitted with a time distribution governed by the lifetime of the muonic atom, as mentioned in \Cref{sec:timedistribution}. For suitable stopping-target materials, this lifetime is substantially longer than the duration of the proton pulse. The signal search is therefore performed in delayed measurement windows that begin after the prompt beam-related activity, or beam flash, has been sufficiently suppressed. In practice, the opening time of the measurement window is determined primarily by the suppression required for pion-induced backgrounds, particularly those arising from radiative pion capture (RPC), as discussed in \Cref{sec:beampion}.

\begin{figure}[tb!]
\centering
	 \includegraphics[width=0.8\textwidth]{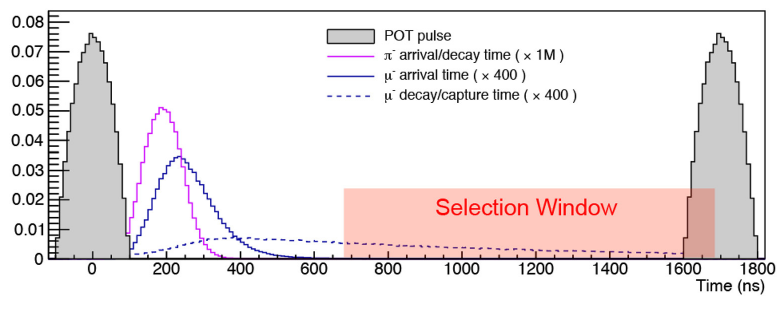}
\caption{Example of the delayed measurement window strategy, shown for the Mu2e timing structure. Proton pulses arrive every $1.695\,\muup\mathrm{s}$, and the arrival-time distributions of pions and muons are indicated. The delayed measurement window, labeled ``Selection window'' in the plot, begins approximately $700\,\mathrm{ns}$ after the proton pulse, when the prompt beam-related activity has been sufficiently suppressed. Reproduced from \citet{Bernstein:2019fyh}, courtesy of the Mu2e Collaboration.}
\label{fig:delayed-time-window}
\end{figure}

As an example of this common delayed measurement window strategy, \Cref{fig:delayed-time-window} shows the timing structure adopted by Mu2e. In Mu2e, the delayed measurement window begins approximately $700\,\mathrm{ns}$ after each proton pulse. Mu2e beamline simulation studies indicate that this delay suppresses the number of pions reaching the stopping target by a factor of $\mathcal{O}(10^{11})$, thereby reducing the RPC background to the required level.\footnote{Some muonic atoms can survive beyond one proton-beam period and may still undergo \mue\ conversion at later times. However, each subsequent proton pulse again produces prompt beam-related backgrounds, so events immediately after each pulse are rejected, and only the corresponding delayed measurement window is used for the signal search.} COMET employs the same delayed measurement window principle in its J-PARC pulsed-beam environment.

The use of a delayed measurement window requires stringent suppression of out-of-time protons between nominal proton pulses. Protons arriving during the measurement window can produce new pions and thereby regenerate prompt beam-related backgrounds. Proton beam extinction, defined as the ratio of the number of out-of-time protons to the number of protons in the pulses, is therefore a central experimental requirement. Next-generation \mue\ conversion searches require extinction at the level of $10^{-10}$ or better.\footnote{This extinction requirement at the $10^{-10}$ level is comparable to the $\mathcal{O}(10^{11})$ pion suppression obtained in Mu2e simulations using a delayed measurement window starting about $700\,\mathrm{ns}$ after the proton pulse.} At the same time, monitoring systems capable of verifying this suppression during operation are needed.

This timing requirement also constrains the choice of stopping-target material. High-$Z$ nuclear targets have shorter muonic-atom lifetimes, as shown in \Cref{fig:muonicatomlifetime}. If the lifetime is too short, too few muonic atoms survive into the delayed measurement window. For experiments employing a delayed measurement window, low-$Z$ nuclear targets are therefore often favored because they provide a useful compromise between signal acceptance and the rejection of beam-related backgrounds. For instance, aluminum, which has a muonic-atom lifetime of $864\,\mathrm{ns}$, is used as the stopping-target material in both COMET and Mu2e.

\subsection{Muon Stopping Target, Detector Requirements, and Signal Selection}

\begin{figure}[tb!]
\centering
	 \includegraphics[width=0.6\textwidth]{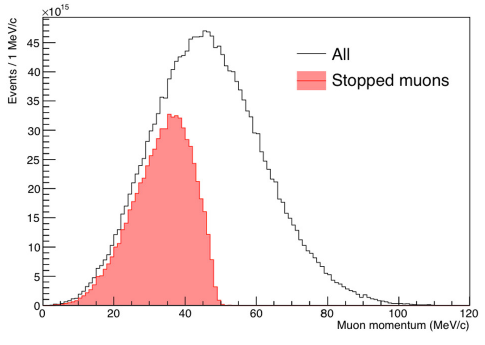}
\caption{Example of muon-stopping-target optimization, shown for the Mu2e configuration. The momentum distributions of muons reaching the stopping target and of those stopping in the target are compared; the stopped-muon distribution is shown in red. Reproduced from \citet{bartoszek2015mu2etechnicaldesignreport}, courtesy of the Mu2e Collaboration.}
\label{fig:stopped-muons}
\end{figure}

Beyond the choice of stopping-target material imposed by timing considerations, the stopping-target geometry should be optimized to maximize the useful stopped-muon yield while preserving the momentum resolution of the outgoing \mue\ conversion electrons. This favors, for example, arrays of thin foils with appropriate spacing between them, which provide sufficient stopping power while minimizing the amount of material traversed by the outgoing signal electrons. \Cref{fig:stopped-muons} illustrates this optimization for the Mu2e configuration by comparing the momentum distributions of muons reaching the stopping target with those of muons stopping within it. The same general considerations apply to COMET, although the detailed momentum acceptance and target geometry differ.

The expected \mue\ conversion signal would appear as a peak at the endpoint of the DIO spectrum, where the spectrum falls steeply.
Therefore, excellent momentum resolution is required for next-generation searches.
At the high rates expected in these experiments, the tracking system must also maintain robust pattern recognition.
Where calorimeters are used, they provide complementary timing, energy, and particle-identification (PID) information.
In particular, PID helps reject cosmic-ray backgrounds in which muons can be misidentified as signal electrons.

Candidate events are selected by requiring the reconstructed track momentum to fall within the signal region, the event time to fall within the delayed measurement window, and the reconstructed track to be geometrically consistent with an origin in the stopping target. Cosmic-ray veto systems provide additional rejection of non-beam backgrounds at the sensitivity levels targeted by next-generation experiments.

A useful measure of the intrinsic statistical reach of an experimental search is the single-event sensitivity (SES), defined as the \mue\ conversion ratio corresponding to one expected signal event. The SES is given by
\begin{equation}
\mathrm{SES} =
\frac{1}{N_{\mu\mathrm{stop}}\, f_{\mathrm{cap}}\, A_{\mu e}},
\end{equation}
where $N_{\mu\mathrm{stop}}$ is the total number of stopped muons, $f_{\mathrm{cap}}$ is the fraction of stopped muons that undergo nuclear capture, and $A_{\mu e}$ is the total signal acceptance. The factor $A_{\mu e}$ includes the detector geometrical acceptance, the acceptances of the momentum and delayed measurement windows, and the trigger and track reconstruction efficiencies. To determine an expected upper limit or discovery sensitivity, the expected background level must also be taken into account.

%
%
%
\section{Current and Future Experiments}
\label{sec:currentfutureexperiments}

Two ongoing next-generation experiments searching for \mue\ conversion are the COMET experiment at the Japan Proton Accelerator Research Complex (J-PARC) in Japan and the Mu2e experiment at Fermi National Accelerator Laboratory (Fermilab) in the United States. Both experiments aim to achieve high stopped-muon rates using solenoid pion-capture systems and curved-solenoid muon-transport systems, as discussed in \Cref{sec:captureandtransport}. These approaches are expected to improve the sensitivity to \mue\ conversion by up to four orders of magnitude relative to that achieved in the previous SINDRUM~II search.

\subsection{Historical Development}

The conceptual basis for next-generation \mue\ conversion experiments was established by the MELC proposal in Russia in 1989~\citep{Dzhilkibaev:1989aa}. MELC introduced a solenoid-based concept for pion capture and muon transport to produce an intense low-energy $\mu^{-}$ beam for \mue\ conversion searches. This concept became the foundation for later experimental designs.

The MECO experiment, proposed at Brookhaven National Laboratory in the United States in 1997, developed this approach into a detailed experimental design~\citep{Molzon:2002sf}. Although MECO was canceled in 2005 after funding was not approved, its design strongly influenced both COMET and Mu2e. After the cancellation of MECO, this experimental approach evolved into two separate programs: Mu2e retained much of the MECO concept, while COMET adopted the same basic strategy with a different experimental layout. Thus, the similarities between COMET and Mu2e reflect their shared design concepts from MELC and MECO, whereas their differences arise from later design choices and from the accelerator environments at J-PARC and Fermilab.

Another related search at J-PARC, the DeeMe experiment, is treated separately because its sensitivity goal and experimental concept differ from those of COMET and Mu2e. A brief summary is given in \Cref{app:deeme}.

\subsection{COMET Program}
\label{sec:comet}

The COMET experiment, whose name stands for COherent Muon to Electron Transition, is designed to search for \mue\ conversion in aluminum at J-PARC in Japan~\citep{COMET:2009qeh}. COMET was originally proposed in 2008 and later reproposed in 2012 as a staged program. COMET Phase-I is designed to perform an intermediate-sensitivity search, validate the beamline, and measure beam-related backgrounds. COMET Phase-II is intended to extend the experimental reach toward the final design sensitivity.

\subsubsection{COMET Phase-I}

COMET Phase-I~\citep{10.1093/ptep/ptz125} aims to achieve a single-event sensitivity of $3.1 \times 10^{-15}$ with approximately 150 days of data taking. Assuming negligible background, this corresponds to an expected 90\,\% confidence-level upper limit of
\begin{equation}
\mathrm{CR}(\mu^{-}\mathrm{Al} \to e^{-}\mathrm{Al}) < 7 \times 10^{-15}.
\end{equation}
This represents an improvement of about two orders of magnitude over the current upper limit set by SINDRUM~II.

COMET Phase-I uses an $8\,\mathrm{GeV}$ proton beam with a beam power of $3.2\,\mathrm{kW}$ from the J-PARC Main Ring. The proton beam impinges on a graphite pion-production target placed inside a $5\,\mathrm{T}$ superconducting pion-capture solenoid. Low-momentum pions emitted backward with respect to the incident proton beam are efficiently captured, and the negative muons produced from their decays are transported to the experimental area.

The transported muons pass through a $90^{\circ}$ curved solenoid before reaching the muon stopping target. This curved transport section provides charge and momentum selection. Additional thin superconducting correction coils attached to the inner side of the curved solenoid coils generate a dipole magnetic field perpendicular to the bending plane. By adjusting this magnetic field, the transmitted particles can be restricted to the desired charge and momentum range. A more detailed description of the curved-solenoid transport principle, including the role of the correction dipole field, is given in \Cref{app:curvedsolenoid}.

The proton beam has a pulsed time structure, with a pulse width of the order of $100\,\mathrm{ns}$ and a repetition period of $1.17\,\muup\mathrm{s}$.\footnote{The RF period of the J-PARC Main Ring is $0.585\,\muup\mathrm{s}$, and protons are filled into every other RF bucket.} This time structure is produced by bunched slow extraction from the J-PARC Main Ring, in which the proton bunch structure in the ring is preserved during extraction, unlike in ordinary slow extraction. Aluminum is used as the muon-stopping-target material because the muonic-atom lifetime in aluminum, about $864\,\mathrm{ns}$, is well matched to this timing. This allows the experiment to open a delayed measurement window starting about $700\,\mathrm{ns}$ after the proton pulse. 

The required proton-beam extinction, of order $10^{-10}$ as discussed in \Cref{sec:timing-extinction}, is achieved primarily by forming an alternating filled-and-empty proton-bunch structure. Residual protons in the nominally empty buckets are further suppressed by appropriate timing of beam extraction from the $3\,\mathrm{GeV}$ Rapid Cycling Synchrotron (RCS) and injection into the J-PARC Main Ring. An extinction level compatible with this requirement has already been demonstrated using the $8\,\mathrm{GeV}$ pulsed proton beam produced by bunched slow extraction at the J-PARC Hadron Experimental Hall~\citep{Noguchi:2022ioy}.

The baseline muon stopping target consists of 17 thin, high-purity aluminum foils, each $200\,\muup\mathrm{m}$ thick and $100\,\mathrm{mm}$ in diameter, arranged with a spacing of $50\,\mathrm{mm}$. This configuration provides sufficient muon-stopping efficiency while minimizing energy loss and multiple scattering of the outgoing \mue\ conversion electrons. The \mue\ conversion ratio is normalized to the total number of stopped negative muons in the aluminum target during the full data-taking period. This number is evaluated by measuring muonic aluminum X-rays with a high-purity germanium detector equipped with appropriate radiation shielding.

For the \mue\ conversion search, the COMET Phase-I detector system includes a cylindrical drift chamber~\citep{Sato:2024zbm} surrounding the aluminum stopping target. The detector system is placed inside a $1\,\mathrm{T}$ superconducting solenoid magnet. The drift chamber has a large inner radius of about $50\,\mathrm{cm}$, which prevents most low-momentum DIO electrons with transverse momenta below approximately $75\,\mathrm{MeV}/c$ from entering the sensitive tracking region and thereby reduces detector occupancy. The 0.5-mm-thick carbon-fiber-reinforced plastic (CFRP) inner wall of the drift chamber serves as an absorber for protons produced by muon capture.

To suppress backgrounds induced by cosmic rays, the detector is surrounded by passive shielding and an active cosmic-ray veto system composed of layers of plastic scintillators and resistive-plate chambers.

In addition to the \mue\ conversion search, COMET Phase-I provides measurements of the muon beam and beam-related backgrounds needed to validate the Phase-II design.

\subsubsection{COMET Phase-II}

\begin{figure}[tb]
\centering
	 \includegraphics[width=\textwidth]{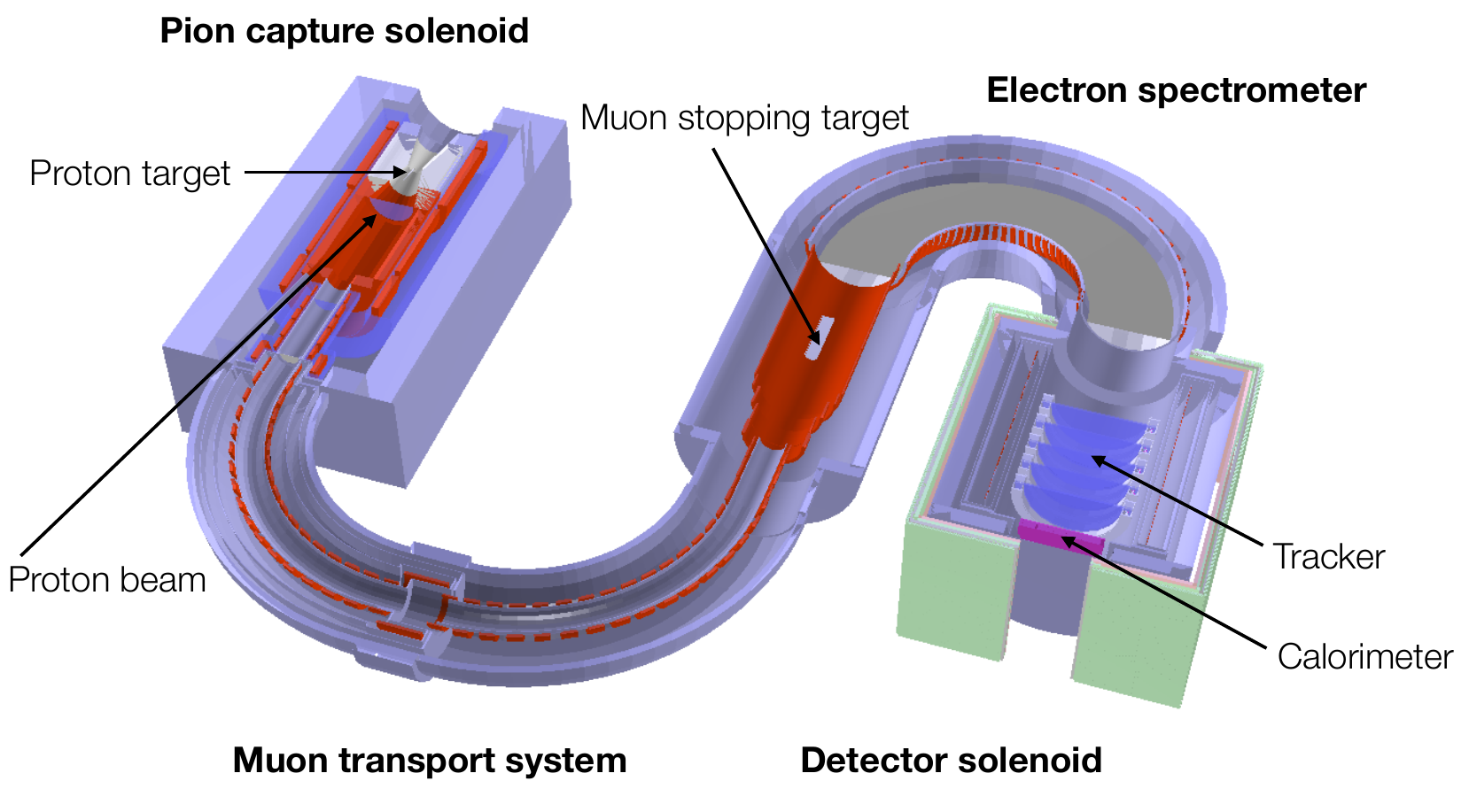}
\caption{Schematic experimental layout of the COMET Phase-II experiment. From left to right: the pion-capture solenoid system; the muon-transport system, consisting of a $180^{\circ}$ curved solenoid; an aluminum muon stopping target; and an electron spectrometer, also consisting of a $180^{\circ}$ curved solenoid, followed by the detector system. A proton beam is injected from the bottom left onto a pion-production target located inside the pion-capture solenoid. Figure courtesy of the COMET Collaboration.}
\label{fig:comet-layout}
\end{figure}

The COMET Phase-II experiment~\citep{Kuno:2013mha,angelique2018cometsubmission2020} is planned to use an $8\,\mathrm{GeV}$ proton beam with a beam power of $56\,\mathrm{kW}$. With approximately 230 days of data taking, COMET Phase-II aims to achieve a single-event sensitivity of $2.6 \times 10^{-17}$, corresponding to an expected 90\,\% confidence-level upper limit of
\begin{equation}
\mathrm{CR}(\mu^{-}\mathrm{Al} \to e^{-}\mathrm{Al}) < 6 \times 10^{-17}.
\end{equation}

The layout of the COMET Phase-II experiment is shown in \Cref{fig:comet-layout}. The solenoid pion-capture system from COMET Phase-I is retained for COMET Phase-II, with additional radiation shielding and a water-cooled tungsten target in the baseline design. The muon-transport system extends the COMET Phase-I design by adding a second $90^{\circ}$ curved solenoid downstream of the first. The two $90^{\circ}$ curved solenoids bend in the same direction, forming a C-shaped $180^{\circ}$ transport section. Each curved solenoid includes an additional dipole coil that produces a correction dipole field. As in COMET Phase-I, this dipole field compensates for the drift of charged particles induced by the curved solenoids and keeps the selected muon beam near the solenoid axis.

In COMET Phase-II, the momentum-dependent displacement (dispersion) accumulates along the longer single-direction C-shaped transport section, enabling more effective selection of low-momentum negative muons before they reach the muon stopping target. The dispersion at the exit is roughly twice that of the $90^{\circ}$ transport section used in COMET Phase-I, improving charge and momentum selection while further reducing pion contamination and other unwanted particles.

The muon stopping target is placed downstream of the muon-transport solenoid in a graded solenoidal magnetic field, where a magnetic mirroring effect is expected, as discussed in \Cref{app:mirroring}. 

A distinctive feature of COMET Phase-II is the additional $180^{\circ}$ curved solenoid installed between the muon stopping target and the detector. This section, which contains a correction dipole field, acts as an electron spectrometer. By adjusting the dipole field inside the curved solenoid, the spectrometer selects electrons with momenta near the \mue\ conversion-electron signal momentum of approximately $105\,\mathrm{MeV}/c$, while suppressing background particles, including low-momentum DIO electrons and positively charged particles such as protons from nuclear muon capture. The curved electron spectrometer also limits the direct line of sight from the muon stopping target to the detector, reducing exposure to neutral particles such as neutrons and photons from the stopping-target region. Together, these effects reduce detector hit rates and provide a cleaner environment for precision momentum measurement.

The detector system consists of a straw-tube gas tracker for measuring the electron momentum and an electromagnetic calorimeter based on LYSO scintillation crystals for measuring the electron energy. Both detectors are operated in vacuum inside a $1\,\mathrm{T}$ magnetic field. These detector technologies are also used for beam diagnostics and performance studies in COMET Phase-I.

Further design optimization studies are being pursued to improve the sensitivity beyond the original Phase-II design goal~\citep{Krikler:2016qij,Yao:2025gaw}.

\subsection{Mu2e Program}

\subsubsection{Mu2e}

The Mu2e experiment~\citep{bartoszek2015mu2etechnicaldesignreport,Bernstein:2019fyh,Mu2e:2022ggl} is a next-generation \mue\ conversion search at Fermilab in the United States. Mu2e was developed at Fermilab following the cancellation of MECO, and its conceptual design was documented in 2012. The experiment uses an aluminum muon stopping target and is designed to accumulate $3.6 \times 10^{20}$ protons on target, corresponding to approximately 680 days of data taking. For the full planned data set, the expected single-event sensitivity is about $3 \times 10^{-17}$, corresponding to an expected 90\,\% confidence-level upper limit of
\begin{equation}
\mathrm{CR}(\mu^{-}\mathrm{Al} \to e^{-}\mathrm{Al}) < 8 \times 10^{-17}.
\end{equation}
Mu2e is also expected to have a $5\,\sigma$ discovery reach at the level of $2 \times 10^{-16}$.

\begin{figure}[tb]
\centering
	 \includegraphics[width=\textwidth]{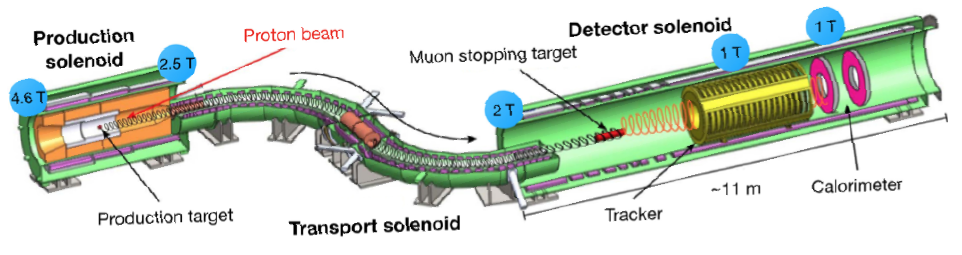}
\caption{Schematic layout of the Mu2e experiment at Fermilab. From left to right, the main components are the Production Solenoid, the Transport Solenoid, and the Detector Solenoid. The magnetic field strengths at different positions are shown. Figure courtesy of the Mu2e Collaboration.}
\label{fig:mu2e-layout}
\end{figure}

The layout of the Mu2e experiment is shown in \Cref{fig:mu2e-layout}. The experiment consists of three main solenoidal sections: the Production Solenoid, the Transport Solenoid, and the Detector Solenoid. An $8\,\mathrm{GeV}$ pulsed proton beam with a beam power of about $8\,\mathrm{kW}$ is delivered from the Fermilab accelerator complex to a radiatively cooled pion-production target. The target is located inside the Production Solenoid, where a graded magnetic field enhances the collection of backward-emitted pions and muons. Muons produced by pion decay in flight are then transported downstream.

The Transport Solenoid consists of two $90^{\circ}$ curved solenoids that bend in opposite directions, forming an S-shaped configuration. The first curved section provides charge and momentum separation, and collimators at its end select negative muons in the desired momentum range. The second curved section bends in the opposite direction, bringing the selected beam back toward the solenoid axis before it enters the Detector Solenoid. This S-shaped configuration preserves the charge and momentum selection provided by the first bend while reducing the net momentum-dependent displacement at the end of the transport section.

The proton beam has a pulsed time structure, with a pulse interval of $1.695\,\muup\mathrm{s}$ and a pulse width of approximately $250\,\mathrm{ns}$. The delayed measurement window begins about $700\,\mathrm{ns}$ after the arrival of the proton beam, as shown in \Cref{fig:delayed-time-window}. This delayed measurement window requires strong suppression of out-of-time protons. Mu2e requires a proton-beam extinction level of $10^{-10}$ or better. This level is achieved by combining the intrinsic extinction provided by beam formation in the Fermilab accelerator complex with an external extinction system based on AC dipoles and collimators. The AC dipoles deflect out-of-time protons into the collimators while transmitting the in-time beam.

The Mu2e aluminum muon stopping target is located in the graded magnetic-field region of the Detector Solenoid and consists of 37 annular aluminum foils spaced by $2.2\,\mathrm{cm}$ along the solenoid axis. Each foil is $105\,\muup\mathrm{m}$ thick, with inner and outer radii of $2.2\,\mathrm{cm}$ and $7.5\,\mathrm{cm}$, respectively. The graded magnetic field in this region helps reflect upstream-going \mue\ conversion electrons back toward the detector acceptance. The number of stopped muons is measured by the stopping target monitor using characteristic X-rays from muonic aluminum atoms, with complementary normalization provided by $\gamma$ rays emitted following nuclear muon capture.

The Mu2e detector system consists of a low-mass straw-tube tracker and an electromagnetic calorimeter, both operated in vacuum inside the $1\,\mathrm{T}$ magnetic field of the Detector Solenoid. The tracker is made of $5\,\mathrm{mm}$-diameter straw tubes with $15\,\muup\mathrm{m}$-thick walls, arranged in an annular, low-mass geometry with a central uninstrumented region. This geometry allows most low-momentum particles, including low-energy DIO electrons and prompt beam-related backgrounds, to pass through without entering the sensitive tracking volume.  The tracker measures \mue\ conversion-electron candidates near $105\,\mathrm{MeV}/c$ with a momentum resolution better than $300\,\mathrm{keV}/c$ FWHM~\citep{Mu2e:2022ggl}. A 0.5-mm-thick polyethylene absorber for protons produced by muon capture is placed between the muon stopping target and the detector. The calorimeter, composed of undoped CsI crystals, provides complementary energy, timing, and position measurements, assists track reconstruction, supports triggering, and helps distinguish electrons from muons.

Antiproton backgrounds are suppressed by placing a thin absorber window in the Transport Solenoid, where it stops slow antiprotons before they can reach the muon stopping target. To suppress cosmic-ray-induced backgrounds, the Detector Solenoid and part of the Transport Solenoid are surrounded by a cosmic-ray veto system made of plastic scintillator counters. The veto system is required to achieve an inefficiency below $10^{-4}$ in order to reduce cosmic-ray-induced backgrounds to an acceptable level.

For the full data set of $3.6 \times 10^{20}$ protons on target, the total expected background is about $0.41$ events. The dominant background categories include cosmic-ray-induced events, DIO electrons, RPC events, antiproton-induced backgrounds, and beam-related processes. Their suppression relies on the combined performance of the cosmic-ray veto, the low-mass tracker, the delayed measurement window, the antiproton absorber, and proton extinction. The detailed breakdown of the expected background contributions is given by \citet{Mu2e:2022ggl}.

\citet{Mu2e:2022ggl} presented an initial data-taking scenario, referred to as Run I, corresponding to about $10\,\%$ of the total planned Mu2e stopped-muon statistics. Under this scenario, Mu2e expects a single-event sensitivity of $2.4\times10^{-16}$, with a total expected background of $0.11\pm0.03$ events. In the absence of a signal, the expected $90\,\%$ C.L. upper limit on the \mue\ conversion ratio is $6.2\times10^{-16}$, corresponding to an improvement of about three orders of magnitude over the SINDRUM~II limit.

\subsubsection{Mu2e-II}

As a future extension of Mu2e, the Mu2e Collaboration has proposed the Mu2e-II program, which aims to improve the sensitivity to \mue\ conversion by at least an order of magnitude~\citep{byrum2022mu2eiimuonelectronconversion}. Mu2e-II is intended to take advantage of the higher proton intensity available from the PIP-II accelerator upgrade at Fermilab. PIP-II, the Proton Improvement Plan II, is a superconducting linear accelerator designed to provide $0.8\,\mathrm{GeV}$ protons with a beam power of $1.6\,\mathrm{MW}$~\citep{Lebedev:2017vnu}. For Mu2e-II, a proton-beam power of about $100\,\mathrm{kW}$ is being considered. Since the proton energy is below the antiproton-production threshold, antiproton-induced backgrounds are not expected to be a limiting issue for Mu2e-II.

A preliminary Mu2e-II sensitivity study assumes an upgraded pion-production target, a reduced-mass straw tracker with thinner straw walls, and a BaF$_2$ calorimeter replacing the undoped CsI calorimeter used by Mu2e. For a five-year running period, corresponding to about $6\times10^{22}$ protons on target and $5.5\times10^{19}$ stopped muons, the estimated single-event sensitivity is $3.25\times10^{-18}$. This corresponds to an improvement of about an order of magnitude relative to the Mu2e sensitivity.

Achieving this sensitivity will require further optimization of the beamline and detector performance under higher-rate conditions. The quoted sensitivity is therefore based on current design assumptions and may be refined as the Mu2e-II design evolves.

\subsection{Comparison between COMET and Mu2e}

COMET and Mu2e share the same basic strategy: solenoid pion capture, curved-solenoid muon transport, pulsed proton beams, delayed measurement windows, aluminum stopping targets, precision tracking, and cosmic-ray vetoing. One main difference lies in the beamline layout downstream of the pion-capture system.

Mu2e uses an S-shaped Transport Solenoid composed of two curved-solenoid sections that bend in opposite directions. The first $90^{\circ}$ curved section provides charge and momentum selection, while the opposite bend partially cancels the dispersion and brings the selected muon beam back toward the solenoid axis before it enters the Detector Solenoid. The annular detector system is placed downstream of the stopping target.

COMET Phase-II uses a C-shaped muon-transport solenoid, in which two $90^{\circ}$ curved solenoids bend in the same direction. 
With correction dipole fields inside the curved solenoids, the central trajectory of the selected muon beam is guided along the curved-solenoid axis, while finite dispersion is retained for momentum selection up to the end of the muon-transport section. In addition, as described above, COMET Phase-II places a dedicated $180^{\circ}$ curved electron spectrometer between the muon stopping target and the detector. This downstream spectrometer provides additional momentum and particle selection before the electrons enter the detector.

Thus, the two experiments implement the same general \mue\ conversion search concept, but with different optimization strategies, particularly in the beam optics of the muon-transport system and in downstream rate suppression.

\subsection{Future Prospects of \texorpdfstring{\mue}{mu-e}\ Conversion Searches}

\subsubsection{Muon Storage Ring Option: PRISM/PRIME}
\label{sec:prism}

Further improvements in the sensitivity of \mue\ conversion searches require not only higher muon intensities but also substantially improved beam quality. In particular, future experiments are expected to suppress pion contamination and reduce the muon momentum spread. The suppression of pions is especially important because pion-induced processes, such as radiative pion capture (RPC) in nuclei, can generate serious beam-related backgrounds.

A muon storage ring provides a possible route to meeting these requirements for an intense, high-purity, low-energy muon beam. One specific realization is the PRISM (Phase-Rotated Intense Slow Muon source) concept, together with the associated PRIME detector system for the \mue\ conversion search~\citep{Kuno:2005mm}. In PRISM, pions produced in the solenoid pion-capture system decay into muons, which are then injected into a fixed-field alternating-gradient (FFA) ring. The FFA ring is designed to accept and store low-momentum muons with the large transverse and longitudinal emittances produced by solenoid pion capture. Further details of the FFA ring and its role in PRISM are summarized in \Cref{app:prism-ffa}.

A major advantage of the storage-ring approach is the strong residual pion suppression. Since muons circulate for several turns before extraction, most residual pions decay along the extended flight path.\footnote{For example, for $40\,\mathrm{MeV}/c$ pions and muons traveling a total flight length of about $200\,\mathrm{m}$ over multiple turns in a storage ring, the pion survival fraction is of order $10^{-39}$, while roughly half of the muons survive.} The resulting small pion contamination would greatly reduce the need for a delayed measurement window to reject pion-related backgrounds. It would also make it easier to use stopping targets with shorter muonic-atom lifetimes, thereby enabling measurements with a wider range of target nuclei, including high-$Z$ nuclei. Such measurements would be valuable for investigating the underlying CLFV mechanism if a signal were observed.

\begin{figure}[htb!]
\centering
	 \includegraphics[width=0.7\textwidth]{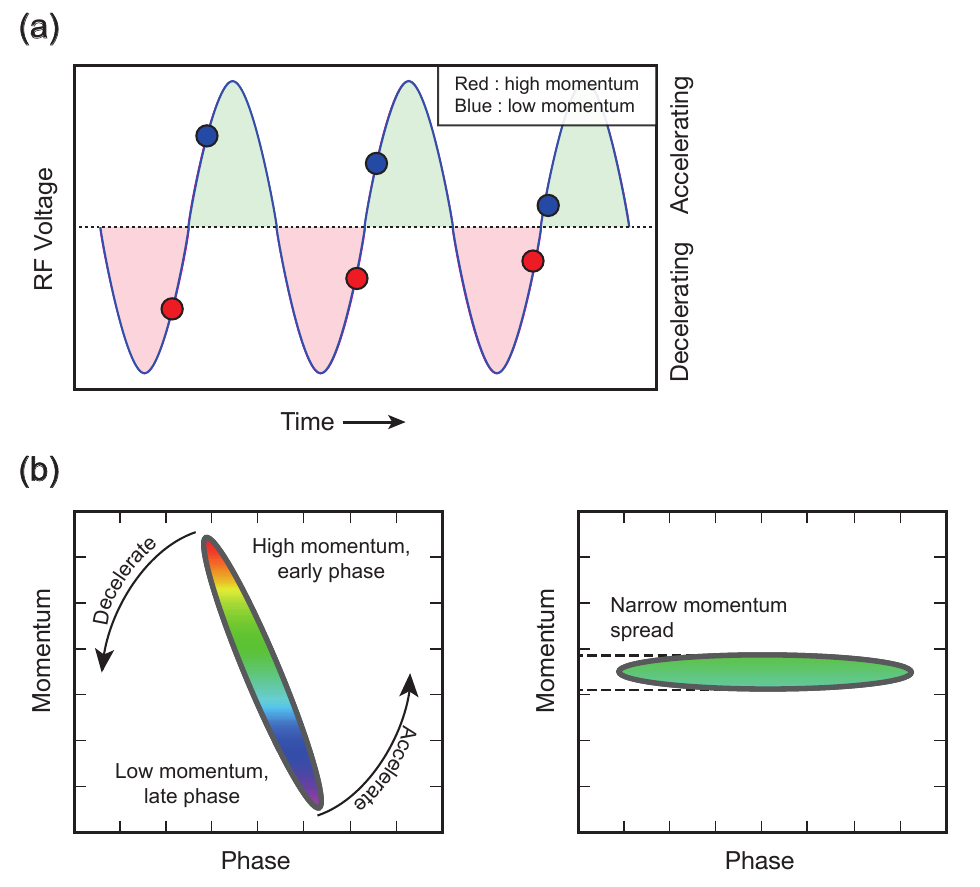}
\caption{Schematic illustration of RF phase rotation. Panel (a) shows the time dependence of the RF voltage. Early-arriving, higher-momentum muons, shown as filled red circles, are decelerated by the RF field, while later-arriving, lower-momentum muons, shown as filled blue circles, are accelerated. After several RF periods, the momentum difference is reduced. Panel (b) shows the corresponding rotation of the beam distribution in longitudinal phase space, represented by the RF phase on the horizontal axis and momentum on the vertical axis. After approximately a quarter-turn rotation, the projection onto the momentum axis becomes narrower, while the distribution in phase becomes broader.}
\label{fig:phase-rotation}
\end{figure}

Another important advantage is phase rotation. In the FFA ring, radio-frequency (RF) electric fields decelerate faster muons and accelerate slower ones, thereby rotating the distribution in longitudinal phase space. Here, the longitudinal phase space is represented by the RF phase, or equivalently by the arrival time, and by the muon momentum, as illustrated in \Cref{fig:phase-rotation}. This technique requires a short proton-beam pulse so that higher-momentum muons arrive earlier than lower-momentum muons. The RF phase is then chosen such that the early-arriving particles are decelerated and the later-arriving particles are accelerated. The resulting narrower momentum spread is expected to improve the stopping efficiency, as suggested by the example distribution for the Mu2e case shown in \Cref{fig:stopped-muons}. It also permits the use of a thinner stopping target, thereby reducing energy loss and multiple scattering of \mue\ conversion electrons, improving the reconstructed momentum resolution, and enhancing DIO-background rejection near the signal region.

The storage-ring scheme also suppresses beam-related electron backgrounds. Beam electrons with momenta near the \mue\ conversion signal momentum, including those produced by pion or muon decays in flight, are not efficiently stored in an FFA ring optimized for low-momentum muons. Moreover, the strong suppression of residual pions 
substantially relaxes the stringent requirement on proton-beam extinction associated with prompt pion-induced backgrounds. Antiprotons are also suppressed. Because of their much larger mass, antiprotons in the accepted momentum range would have much lower velocities than muons and therefore different revolution times and RF phases. As a result, they would not be expected to remain synchronized with the phase-rotated muon bunch or to be efficiently extracted from the FFA ring.

PRIME is an experimental spectrometer concept proposed for use with the PRISM beam and is designed to provide strong suppression of backgrounds associated with muons stopped at a high rate in the target. This is achieved using a curved-solenoid spectrometer equipped with correction dipole fields and a bending angle exceeding $360^{\circ}$, which provides a large separation between signal electrons and low-energy backgrounds.

\begin{figure}[htb!]
\centering
	 \includegraphics[width=0.6\textwidth]{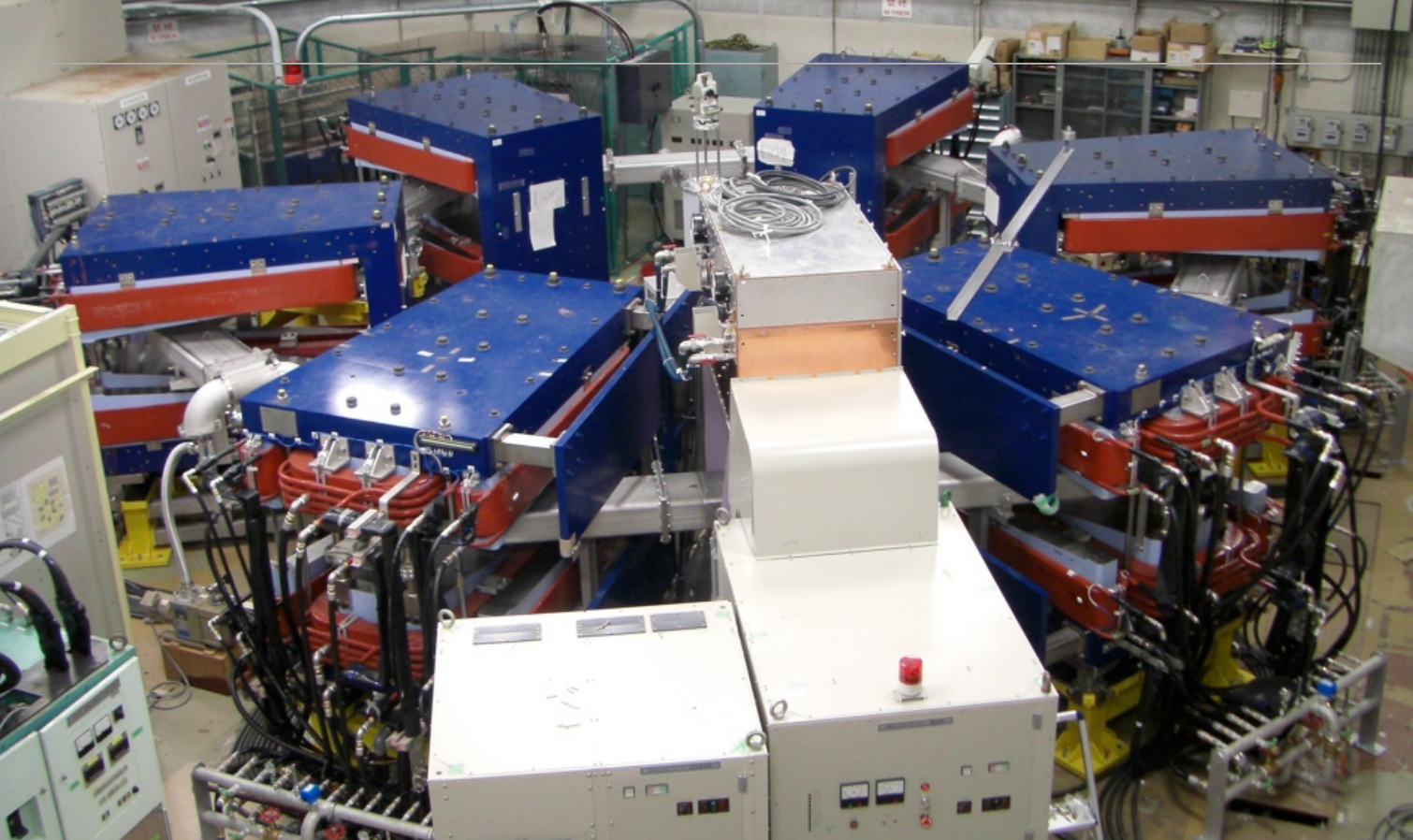}
\caption{Prototype six-sector PRISM-FFA ring constructed at the Research Center for Nuclear Physics (RCNP), The University of Osaka, to demonstrate phase rotation. The ring consisted of six FFA magnet cells, and the white device in the foreground is the RF cavity. Photograph courtesy of the Research Center for Nuclear Physics, The University of Osaka.}
\label{fig:prototype-prism}
\end{figure}

The phase-rotation concept was tested using a prototype PRISM-FFA ring assembled in 2007 at the Research Center for Nuclear Physics (RCNP), The University of Osaka. The prototype ring, shown in \Cref{fig:prototype-prism}, consisted of six FFA magnet cells and an RF cavity. Phase rotation was demonstrated using $\alpha$ particles under conditions chosen to reproduce the essential features of muon phase rotation. The results were reported in 2010~\citep{Sato:2008zze,Sato:2010wepe044}. This test provided an important proof of principle for the PRISM approach. Further details are given in \Cref{app:prism-ffa-prototype}.

The PRISM/PRIME approach therefore represents a qualitatively different strategy from those of COMET and Mu2e. By using a storage ring to improve the purity and reduce the momentum spread of the muon beam before it reaches the muon stopping target, it offers a route to substantially improved sensitivity in future \mue\ conversion searches.

\subsubsection{Advanced Muon Facility at Fermilab}

The Advanced Muon Facility (AMF) is a proposed high-intensity muon science complex at Fermilab~\citep{Corrodi:2023mza,Middleton:2025ici}. Building on the high proton-beam power enabled by PIP-II, the AMF concept aims to exploit proton beams with power approaching the megawatt scale to establish an intense muon-beam facility. It is intended to provide very intense $\mu^+$ and $\mu^-$ beams for a broad muon-physics program. A central goal would be a flagship CLFV program covering three representative muon CLFV processes: \muegamma\ decay, \mueee\ decay, and \mue\ conversion in nuclei. Such a program could achieve sensitivities orders of magnitude beyond those of currently planned experiments or provide complementary measurements if a CLFV signal is observed in the upcoming generation of experiments.

For \mue\ conversion, AMF would build on the FFA storage-ring strategy described above. A compressed proton bunch would strike a pion-production target inside a pion-capture solenoid, and the resulting muon beam would be injected into an FFA ring. As discussed above, phase rotation would reduce the momentum spread, while storage in the ring would suppress residual pion contamination. The resulting high-purity muon beam could substantially reduce the need for a delayed measurement window and make the use of high-$Z$ muon stopping targets, such as gold, experimentally more attractive.

The ability to control the extracted muon momentum is another important feature of the AMF concept. A relatively low central momentum in the range of $20$--$40\,\mathrm{MeV}/c$, for example, would allow efficient stopping in thinner targets.

The AMF concept remains at an early design stage and would require substantial accelerator and target-system R\&D. Key issues include proton-bunch compression, operation of a high-power pion-production target inside a pion-capture solenoid, the design of a suitable FFA lattice, and efficient injection and extraction of both positive and negative muon beams. The R\&D program also has potential synergies with future muon-collider studies, particularly in areas such as high-power targetry, pion-capture solenoids, and muon-beam manipulation. In this sense, AMF represents a long-term facility concept for extending the reach of \mue\ conversion searches beyond the upcoming generation of experiments and their near-term upgrades, while also supporting a broader muon-physics program.

%
%
%
\section{Complementary Search: \mup\ Conversion in Nuclei}

\subsection{Physics Motivation}

A process complementary to \mue\ conversion that can be searched for using similar experimental setups is \mup\ conversion in nuclei. It is described by
\begin{equation}
\mu^{-} + N(A,Z) \to e^{+} + N(A,Z-2),
\label{eq:muep}
\end{equation}
where $N(A,Z)$ denotes the initial nucleus with mass number $A$ and atomic number $Z$, while $N(A,Z-2)$ denotes the final nucleus, which may be either in its ground state or in an excited state. This process changes the total lepton number $L$ by two units, $|\Delta L|=2$, and also changes both the muon and electron flavor numbers, with $\Delta L_{\mu}=\Delta L_e=-1$. Here, $L_{\mu}$ and $L_e$ are the muon and electron flavor numbers, respectively. Its observation would therefore provide evidence for lepton-number violation (LNV) and CLFV. 

At the nucleon level, this process requires the participation of at least two protons:
\begin{equation}
\mu^{-} + p + p \to e^{+} + n + n .
\label{eq:muep-elementary}
\end{equation}
Thus, unlike coherent \mue\ conversion, the final nucleus differs from the initial one, and no coherent enhancement is expected, even for transitions to the nuclear ground state.

Early theoretical studies considered \mup\ conversion in nuclei primarily as a probe of Majorana-neutrino exchange~\citep{Kamal:1979vw}. Subsequent nuclear-structure calculations investigated mechanisms mediated by light and heavy Majorana neutrinos in $^{48}\mathrm{Ti}$ and showed that the corresponding rates are extremely small under existing constraints on neutrino parameters~\citep{Domin:2004tk}.  More recent studies based on effective field theory have instead examined \mup\ conversion as a probe of general high-scale $|\Delta L|=2$ new physics, including short-range contributions induced by higher-dimensional effective operators~\citep{Geib:2016atx,Geib:2016daa,Berryman:2016slh}.

This process therefore provides a probe of $|\Delta L|=2$ physics, complementary to neutrinoless double-beta decay and collider searches. A recent review of experimental searches for  \mup\ conversion is available~\citep{Lee:2021hnx}.

\subsection{Experimental Signature of \mup\ Conversion}

For a transition to the nuclear ground state, the experimental signature is a single, approximately monoenergetic positron with energy $E_{\mu e^+}$ given by
\begin{equation}
E_{\mu e^+}
\approx
m_\mu - B_\mu - E_{\mathrm{recoil}}
+ M(A,Z) - M(A,Z-2).
\label{eq:mup-signal}
\end{equation}
Here, $m_\mu$ is the muon mass, $B_\mu$ is the binding energy of the $1S$ state of the muonic atom, and $M(A,Z)$ and $M(A,Z-2)$ denote the nuclear masses of the nuclei $N(A,Z)$ and $N(A,Z-2)$, respectively. The recoil energy of the final-state nucleus can be approximated as $E_{\mathrm{recoil}} \approx [m_\mu - B_\mu + M(A,Z) - M(A,Z-2)]^2/(2M(A,Z-2))$. 
For an aluminum target, for example, $E_{\mu e^+}=92.3\,\mathrm{MeV}$, which is lower than the electron energy of $104.97\,\mathrm{MeV}$ for \mue\ conversion in aluminum.

Transitions to excited nuclear states can also be important and, depending on the underlying interaction and nuclear structure, may have rates comparable to those of the ground-state transition. In such cases, the positron signal energy is reduced by the excitation energy, and the signal region can be broadened when many excited states cannot be resolved.

The conversion ratio (CR) is conventionally normalized to the total ordinary muon capture rate in the same nucleus, $\Gamma_{\mathrm{cap}}$, and is given by
\begin{equation}
\mathrm{CR}(\mu^{-}N \rightarrow e^{+}N^{\prime})
\equiv
\frac{\Gamma(\mu^{-}N \rightarrow e^{+}N^{\prime})}
{\Gamma_{\mathrm{cap}}}.
\label{eq:mup-conversion-rate}
\end{equation}

\subsection{Backgrounds and Target Selection}

Potentially dominant backgrounds in \mup\ conversion searches arise from radiative muon capture (RMC) and radiative pion capture (RPC). As discussed in \Cref{sec:rmc-background}, the RMC background arises from
\begin{equation}
\mu^{-} + N(A,Z) \to \nu_{\mu} + N(A,Z-1) + \gamma ,
\quad \gamma \to e^- e^+ .
\label{eq:rmc-lnv}
\end{equation}
In RMC, a real photon can convert externally in material into an $e^-e^+$ pair, while an off-shell photon can internally convert into an $e^-e^+$ pair. If the positron has sufficiently high energy, it can mimic the \mup\ conversion signal. For the case in which the final nucleus remains in its ground state and no nucleons are emitted, the maximum positron energy can be approximated by the RMC photon endpoint, $E_{\mathrm{RMC}}^{\mathrm{max}}$, when the electron mass and small differences in recoil energy are neglected:
\begin{equation}
E_{\mathrm{RMC}}^{\mathrm{max}}
\approx
m_\mu - B_\mu - E_{\mathrm{recoil}}
+ M(A,Z) - M(A,Z-1).
\label{eq:rmc-background}
\end{equation}
A useful target-selection criterion is therefore
\begin{equation}
E_{\mu e^+} > E_{\mathrm{RMC}}^{\mathrm{max}} ,
\end{equation}
which leads to the nuclear-mass condition~\citep{Yeo:2017fej}
\begin{equation}
M(A,Z-1) > M(A,Z-2) .
\label{eq:mass}
\end{equation}
Only a limited set of target nuclei satisfy this condition while also having muonic-atom lifetimes suitable for a delayed measurement window. The target choice is therefore more constrained than in \mue\ conversion searches.

In addition, the RMC spectrum near its endpoint and the distribution of nuclear final states, including possible neutron or proton emission, are not yet fully measured. This makes the RMC background particularly challenging, because even a small, poorly constrained high-energy tail can affect the sensitivity of the search.

RPC backgrounds can be suppressed by using a delayed measurement window and strong beam extinction, as in \mue\ conversion searches.

\subsection{Past Experimental Searches and Future Prospects}

\begin{table}[tb]
\centering
\caption{Upper limits on \mup\ conversion from past experiments. Ground-state and excited-state transitions are shown separately where available. The search marked with $\dagger$ used radiochemical detection of the $^{127}\mathrm{Sb}$ activity produced after irradiating an $^{127}\mathrm{I}$ target with negative muons.}
\vspace{5mm}
\begin{tabular}{lccccl} \hline
Year & Target & Facility & \multicolumn{2}{c}{90\,\% C.L. upper limits}  & References \\ 
& & & Ground state & Excited state & \\ \hline\hline
1962 & Cu & CERN & \multicolumn{2}{c}{$2.2 \times 10^{-7}$} &  \cite{Conforto:1962zz} \\
1972 & Cu & SREL & \multicolumn{2}{c}{$2.6 \times 10^{-8}$} &  \cite{PhysRevLett.28.1469} \\ 
1978 & $^{32}\mathrm{S}$ & SIN & \multicolumn{2}{c}{$1.5 \times 10^{-9}$} & \cite{Badertscher:1978mu} \\
1980 & $^{32}\mathrm{S}$ & SIN & \multicolumn{2}{c}{$9 \times 10^{-10}$}  &  \cite{Badertscher:1980bt}  \\ 
1980 & $^{127}\mathrm{I}$ & $-$ & \multicolumn{2}{c}{$3 \times 10^{-10}$}  &  \cite{Abela:1980rs}$^{\dagger}$ \\ 
1988 & Ti & TRIUMF & \multicolumn{2}{c}{$1.7 \times 10^{-10}$}  & \cite{Ahmad:1988ur}  \\ 
1993 & Ti & PSI & $4.3 \times 10^{-12}$  & $8.9 \times 10^{-11}$  & \cite{DOHMEN1993631} \\ 
1998 & Ti & PSI & $1.7 \times 10^{-12}$ & $3.6 \times 10^{-11}$ & \cite{SINDRUMII:1998mwd} \\ \hline
\end{tabular}
\label{tb:muplist}
\end{table}

Past searches for \mup\ conversion are summarized in \Cref{tb:muplist}. The most stringent limits were obtained by the SINDRUM II experiment at PSI using a titanium target~\citep{SINDRUMII:1998mwd}. The experiment searched for
$\mu^- \mathrm{Ti} \to e^+ \mathrm{Ca}$ and obtained
\begin{equation}
\mathrm{CR}(\mu^- \mathrm{Ti} \to e^+ \mathrm{Ca}~(\mathrm{GS}))
< 1.7 \times 10^{-12},
\end{equation}
for the ground-state transition, and
\begin{equation}
\mathrm{CR}(\mu^- \mathrm{Ti} \to e^+ \mathrm{Ca}~(\mathrm{EX}))
< 3.6 \times 10^{-11},
\end{equation}
for transitions to excited states, assuming excitation through the giant dipole resonance. 

Next-generation \mue\ conversion experiments could also enable searches for \mup\ conversion with sensitivities approaching $\mathcal{O}(10^{-16})$, subject to experimental acceptance and RMC-background control. Such searches would be highly complementary to \mue\ conversion, since they probe LNV in addition to CLFV.

%
%
%
\section{Outlook}

Charged lepton flavor violation (CLFV) remains one of the most sensitive probes of physics beyond the Standard Model. In the Standard Model with neutrino masses and mixing, CLFV rates are far below any anticipated experimental sensitivity. The observation of a CLFV process would therefore provide unambiguous evidence for new physics. Among the available muon CLFV processes, \mue\ conversion in nuclei offers a particularly powerful approach because it provides a clean monoenergetic-electron signature and can make effective use of large stopped-muon samples without relying on coincidence measurements among multiple final-state particles.

The next generation of \mue\ conversion searches will significantly extend the experimental reach. COMET and Mu2e are designed to improve sensitivity by approximately four orders of magnitude over previous experiments. This improvement will be achieved primarily through large samples of low-energy stopped muons produced using solenoid pion capture and curved-solenoid muon transport, together with pulsed proton beams, delayed measurement windows, and stringent proton-beam extinction to suppress beam-related backgrounds.

Beyond COMET and Mu2e, further progress will require both higher muon intensity and improved beam quality. Mu2e-II at Fermilab aims to exploit the higher proton-beam power available from PIP-II, while PRISM/PRIME and AMF-type concepts pursue a qualitatively different regime based on muon-beam manipulation using a storage ring. In particular, strong suppression of residual pions could allow the use of higher-$Z$ muon stopping targets with shorter muonic-atom lifetimes, while a reduced muon momentum spread could permit thinner stopping targets and thereby improve both momentum resolution and signal acceptance. Such developments could extend \mue\ conversion searches toward sensitivities at or below the $10^{-18}$ level and improve their power to discriminate among theoretical models.

If a signal is observed, measurements using multiple target nuclei and comparisons among multiple CLFV processes will be essential for identifying the underlying interaction.
Because overlap integrals and nuclear response functions depend on the target, different nuclei are sensitive to different combinations of CLFV operators. Comparisons of \mue\ conversion with \muegamma\ and \mueee\ decays could help distinguish four-fermion contact interactions from dipole-dominated contributions. Complementary searches for \mup\ conversion would provide additional sensitivity to lepton-number violation (LNV) in the charged-lepton sector.

Continued progress will require improved background control and nuclear-physics inputs. 
From an experimental perspective, a precise understanding of muon decay in orbit (DIO), radiative muon capture (RMC), radiative pion capture (RPC), and antiproton production rates is needed to characterize background sources, as well as effective cosmic-ray background rejection techniques.
At the same time, improved knowledge of nuclear charge, proton, and neutron distributions will be needed to refine overlap-integral calculations. Analyses based on nonrelativistic effective theory (NRET) will also require reliable calculations of the relevant nuclear response functions and matrix elements for different CLFV operators. These inputs will be essential for interpreting any observed signal in terms of specific CLFV operators.

The coming generation of experiments represents both a major discovery opportunity and the foundation of a broader future program. Whether or not COMET and Mu2e observe a signal, their results will strongly constrain models of new physics and guide the design of subsequent facilities. With continued experimental and theoretical progress, \mue\ conversion searches will remain a central component of the CLFV program.

%

\begin{ack}[Acknowledgments]%

The author thanks Sacha Davidson and Robert Bernstein for their careful
reading of the manuscript and valuable comments. He acknowledges support from
the Munich Institute for Astro-, Particle and BioPhysics (MIAPbP), which is
funded by the Deutsche Forschungsgemeinschaft (DFG, German Research
Foundation) under Germany's Excellence Strategy --EXC 2094-- 390783311 and
by the Technical University of Munich (TUM), and is grateful for the
opportunity to participate in the MIAPbP ``Precision Lepton Physics''
workshop.

\end{ack}


\appendix
\section{Solenoid Beam Optics: Solenoidal Capture and Transport of Muon Beams}
\label{app:capture_transport}

Next-generation \mue\ conversion searches require the efficient production and transport of intense beams of low-momentum negative muons to achieve high sensitivity. The muons must be transported to and stopped in a muon stopping target, while beam-related backgrounds must be strongly suppressed. To meet these requirements, COMET and Mu2e employ beamlines based on superconducting solenoids. Their basic elements are a pion-capture system, which collects pions produced at a pion-production target over a large solid angle in a high solenoidal field, and a muon-transport system, which selects low-momentum negative muons from pion decay and guides them to the muon stopping target. This appendix summarizes the basic beam optics of these systems.

\subsection{Pion Capture Using a Solenoidal Magnetic Field}
\label{app:pioncapture}

A high-intensity muon source begins with the production and capture of charged pions. By placing a pion-production target inside a strong solenoidal magnetic field, pions produced within the accepted range of transverse momentum can be confined to helical trajectories and guided into the downstream beamline. This geometry provides a large solid-angle acceptance, even for pions emitted at large angles relative to the beamline axis; see \Cref{fig:pion-capture-solenoid}.

The dimensions and material of the pion-production target must be optimized to maximize the useful low-momentum pion yield while limiting pion reabsorption and managing beam-induced heating. In general, a long, small-radius cylindrical target is favored: its length is chosen to provide a high probability of proton interactions, typically corresponding to about one nuclear interaction length, while its radius must be small enough to allow the produced pions to escape before being reabsorbed. The optimum geometry depends on the specific experimental arrangement. High-$Z$ materials are generally favored because they provide high pion-production yields. At the same time, the target must withstand large heat loads and radiation damage, requiring a material with thermal and mechanical stability, radiation tolerance, and compatibility with the target-cooling scheme. For this reason, materials such as graphite are often chosen when thermal and radiation constraints dominate. 

The pion-capture system consists of the pion-production target surrounded by a large-aperture superconducting solenoid. Substantial shielding is required between the target region and the superconducting coils to limit radiation damage and heat loads in the magnets. The magnet design must also incorporate radiation-tolerant materials suitable for operation near the pion-production target, such as aluminum-stabilized superconductors and radiation-resistant resin for the coils.

For a pion produced near the solenoid axis, the maximum transverse momentum that can be transported without striking the limiting aperture of the pion-capture system, $p_{T}^{\mathrm{max}}$, in $\mathrm{GeV}/c$, is approximately
\begin{equation}
p_{T}^{\mathrm{max}}
\simeq
0.3\,B\,
\left( \frac{R_{\mathrm{cap}}}{2} \right),
\end{equation}
where $B$ is the solenoidal magnetic-field strength in tesla and $R_{\mathrm{cap}}$ is the effective open-aperture radius for pion capture in meters. In practical capture-solenoid designs, $R_{\mathrm{cap}}$ is often set by the inner bore of the radiation shielding surrounding the production target rather than by the superconducting-coil bore itself. A stronger magnetic field and a larger open aperture therefore increase the transverse-momentum acceptance of the pion-capture system. For example, for $B=5\,\mathrm{T}$ and $R_{\mathrm{cap}}=0.20\,\mathrm{m}$, the transverse-momentum acceptance is $p_T^{\mathrm{max}}\simeq 0.15\,\mathrm{GeV}/c$.

Since $p_T^{\mathrm{max}}$ is proportional to $B R_{\mathrm{cap}}$, the accepted area in transverse-momentum space increases approximately as $(B R_{\mathrm{cap}})^2$, although the useful pion yield also depends on the details of the beamline design.  COMET and Mu2e employ high-field pion-capture solenoids with peak fields exceeding $4\,\mathrm{T}$. In both designs, the pion-capture region efficiently collects low-momentum pions and muons produced by pion decay for downstream transport~\citep{10.1093/ptep/ptz125,bartoszek2015mu2etechnicaldesignreport}.

\subsection{Magnetic Mirroring in a Graded Solenoidal Field}
\label{app:mirroring}

Particle acceptance can be increased by using a graded solenoidal magnetic field near either a pion-production target or a muon stopping target. If the magnetic-field strength increases in the direction opposite to the desired particle transport, particles initially emitted in that direction can be reflected by the magnetic-mirror effect and redirected toward the useful acceptance. Near a pion-production target, this effect can enhance the collection of secondary pions and decay muons. Near a muon stopping target, the same principle can increase the acceptance for \mue\ conversion electrons initially emitted away from the detector region.

Let $\theta_0$ denote the initial pitch angle between the particle momentum and the magnetic-field direction at its point of origin. In the adiabatic approximation, reflection occurs when
\begin{equation}
\sin^2\theta_0 \geq \frac{B_0}{B_{\mathrm{max}}},
\end{equation}
where $B_0$ is the magnetic-field strength at the particle's point of origin and $B_{\mathrm{max}}$ is the maximum field encountered along its trajectory. This magnetic-mirroring principle is used around the pion-production target in Mu2e and around the muon stopping target in both Mu2e and COMET Phase-II.

\subsection{Adiabatic Transformation into a More Parallel Beam}

Pions captured in the solenoidal magnetic field are produced over a broad angular range and may initially have large pitch angles relative to the beamline axis. As a result, their transverse momenta can be large, and their helical trajectories may be poorly matched to downstream transport. It is therefore advantageous to transform the captured beam into a more forward-directed beam.

\begin{figure}[tb]
\centering
	 \includegraphics[width=0.4\textwidth]{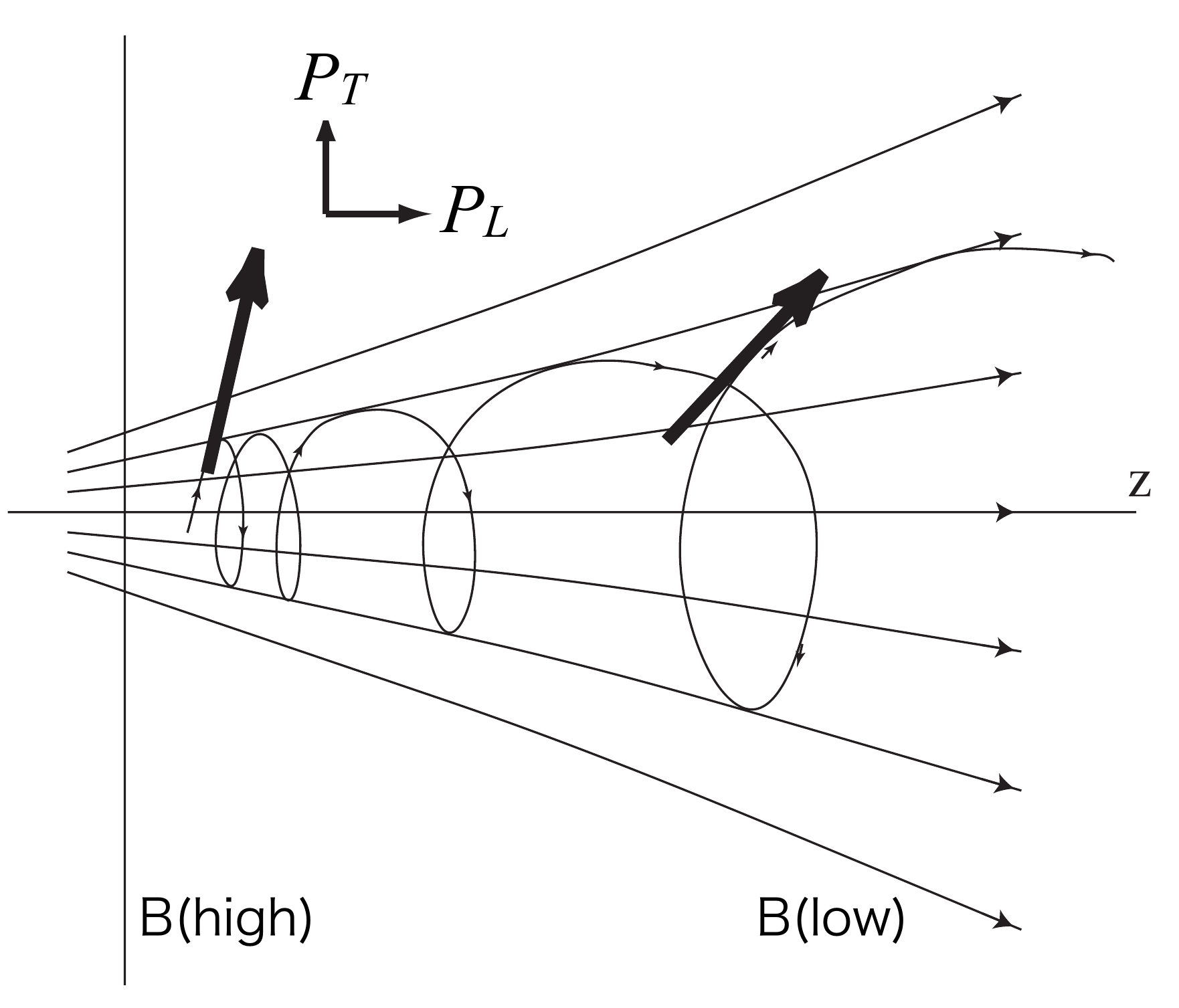}
\caption{Schematic illustration of adiabatic beam transformation. As the solenoidal magnetic-field strength gradually decreases, the transverse momentum of the captured particles decreases and the beam becomes more parallel to the beamline axis, while its transverse size increases.}
\label{fig:adiabatic_transfer}
\end{figure}

This transformation is achieved by gradually decreasing the solenoidal magnetic-field strength downstream of the capture region. When the magnetic field changes adiabatically, the quantity $p_T^2/B$ is approximately conserved.\footnote{This relation follows from conservation of the first adiabatic invariant, commonly referred to as the magnetic moment in plasma and beam physics, for a charged particle moving in a magnetic field that varies slowly over the scale of its cyclotron motion.} Since the helical-orbit radius $r$ is proportional to $p_T/B$, this invariant can also be expressed in terms of the product of $r$ and $p_T$:
\begin{equation}
 \frac{p_T^2}{B} \propto p_T r = \mathrm{constant}.
\end{equation}
Thus, as the magnetic field decreases, the transverse momentum $p_T$ decreases while the helical-orbit radius $r$ increases. Since the total momentum remains unchanged in a static magnetic field, the reduction in transverse momentum corresponds to an increase in the longitudinal component $p_L$. The resulting beam is therefore more closely aligned with the beamline axis, at the cost of an increased transverse beam size. This adiabatic transformation is illustrated in \Cref{fig:adiabatic_transfer}.

For example, if the field strength is reduced by a factor of two, the adiabatic relation gives $p_T \to p_T/\sqrt{2}$ and
$r \to \sqrt{2}\,r$. The reduced angular spread improves the transmission of pions and their decay muons into the downstream transport beamline.

This adiabatic transformation is employed in both COMET and Mu2e. In COMET, the pion-capture field decreases from approximately $5\,\mathrm{T}$ near the production target to $3\,\mathrm{T}$ toward the
entrance of the transport solenoid. In Mu2e, the magnetic field in the Production Solenoid decreases from approximately
$4.6\,\mathrm{T}$ near the production target to $2.5\,\mathrm{T}$ toward
the entrance of the Transport Solenoid~\citep{10.1093/ptep/ptz125,bartoszek2015mu2etechnicaldesignreport}.

\subsection{Charge and Momentum Selection Using a Curved Solenoid}
\label{app:curvedsolenoid}

Following pion capture and decay, the muon beam must be transported to the stopping target while particles with undesired electric charge or momentum are suppressed. This selection can be achieved using a curved-solenoid geometry, which preserves the large acceptance of solenoidal transport while introducing a charge- and momentum-dependent transverse displacement of the beam, or dispersion.

In a straight solenoidal magnetic field, a charged particle follows a helical trajectory about the solenoid axis. In a curved solenoid, the center of this helical motion drifts in a direction perpendicular to the plane of curvature. The drift displacement $D$, in meters, is given by
\begin{align}
D &=
\frac{1}{0.3\,qB}
\left(\frac{s}{R}\right)
\frac{p_L^2+\frac{1}{2}p_T^2}{p_L}
\nonumber \\
&=
\frac{1}{0.3\,qB}
\left(\frac{s}{R}\right)
\frac{p}{2}
\left(
\cos\theta+\frac{1}{\cos\theta}
\right),
\label{eq:curvedsolenoiddrift}
\end{align}
where $q$ is the particle charge in units of the elementary charge, including its sign; $B$ is the
solenoidal magnetic-field strength on the reference axis, in tesla; $s$ and $R$
are the arc length and radius of curvature of the curved solenoid,
respectively, in meters; and $p_L=p\cos\theta$ and $p_T=p\sin\theta$ are the
longitudinal and transverse momentum components, respectively, in $\mathrm{GeV}/c$, with $\theta$ denoting the pitch angle. The ratio $s/R=\theta_{\mathrm{bend}}$ is the total bending angle of the solenoid.

The direction of the drift is determined by the sign of the particle's electric charge, the orientation of the solenoidal magnetic field relative to the beam direction, and the direction of the solenoid curvature. In particular, the drift direction reverses when the sign of the particle charge is reversed, while its magnitude generally increases with the particle momentum and the solenoid bending angle. Consequently, positively and negatively charged particles drift in opposite directions, while particles with different momenta undergo different transverse displacements. Collimators placed in a region of large dispersion can therefore select the desired charge and momentum range.

\begin{figure}[tb]
\centering
	 \includegraphics[width=0.6\textwidth]{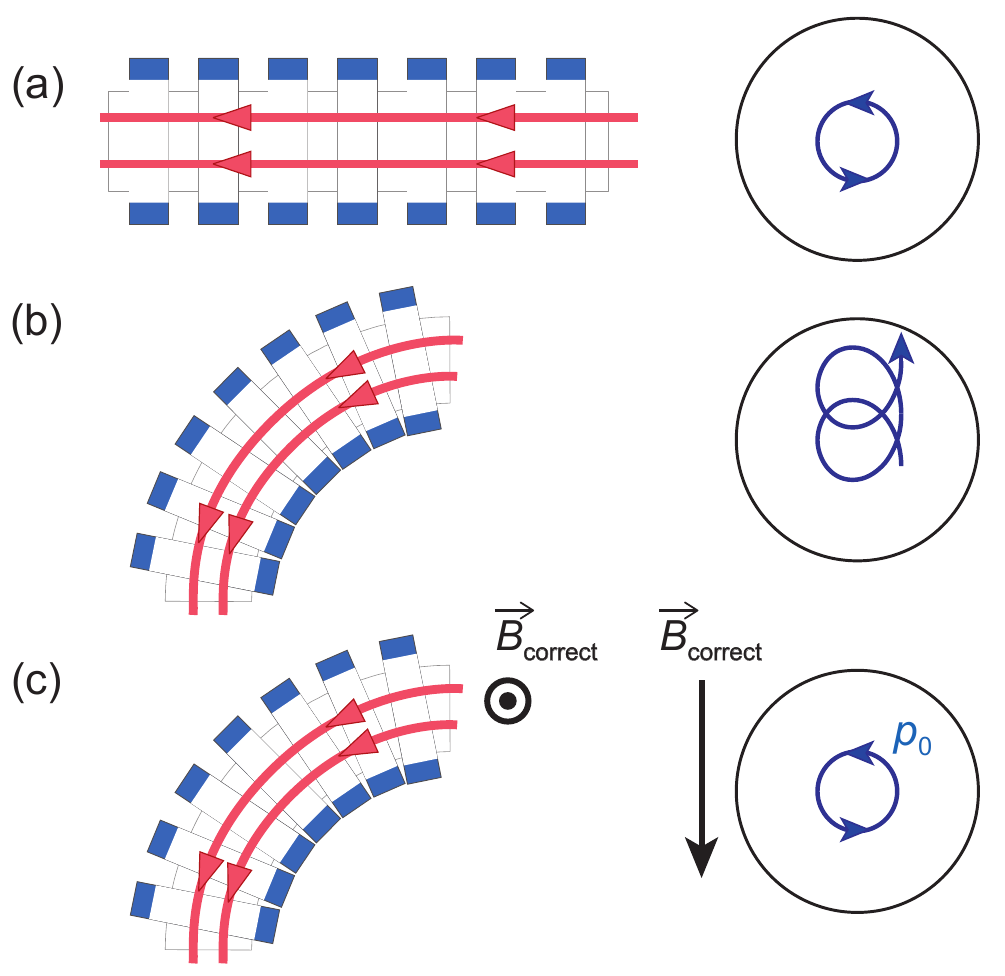}
\caption{Schematic illustration of particle transport in a straight and a
curved solenoid. In the curved section, the center of the helical trajectory
drifts perpendicular to the plane of curvature, with a displacement that
depends on the particle charge and momentum. A correction dipole field,
chosen for a reference momentum $p_0$, keeps the helical-trajectory centers
of the desired particles near the solenoid axis, while particles with
undesired charge or momentum remain displaced and can be rejected by
collimators.}
\label{fig:curvedsolenoid}
\end{figure}

To keep the centers of the helical trajectories of muons with a selected reference momentum $p_0$ close to the solenoid axis, a correction dipole field, $B_{\mathrm{correct}}$, may be superimposed on the solenoidal field. This dipole field is applied perpendicular to the bending plane of the curved solenoid. Its magnitude and direction are chosen so that the transverse drift of the reference particles is compensated.\footnote{Once the drift direction and the polarity of the longitudinal solenoidal field relative to the beam direction are specified, the required direction of the correction dipole field is uniquely determined.} The required correction field is given by
\begin{equation}
B_{\mathrm{correct}}
=
\frac{1}{0.3\,qR}
\frac{p_0}{2}
\left(
\cos\theta_0+\frac{1}{\cos\theta_0}
\right),
\label{eq:curvedsolenoidcorrection}
\end{equation}
where $B_{\mathrm{correct}}$ is in tesla, $p_0$ is in $\mathrm{GeV}/c$, and $\theta_0$ is the pitch angle of the selected trajectory. Muons with the selected charge and momentum are thereby transported with their helical-trajectory centers close to the solenoid axis, whereas particles with undesired charge or momentum remain displaced from the solenoid axis and can be removed by collimators or intercepted by the beamline aperture; see \Cref{fig:curvedsolenoid}.

Since the drift displacement is proportional to the total bending angle, increasing $\theta_{\mathrm{bend}}$ improves the spatial separation of particles with different momenta. By contrast, varying the solenoidal field strength alone provides limited improvement in separation because both the drift displacement and the helical-orbit radius scale inversely with $B$. Curved solenoids with sufficiently large bending angles are therefore effective elements for simultaneous charge and momentum selection in intense muon beamlines~\citep{10.1093/ptep/ptz125}.

\subsection{Demonstration of Pion Capture and Muon Transport: The MuSIC Facility}
\label{app:music}

The MuSIC facility at the Research Center for Nuclear Physics (RCNP), The University of Osaka, provided an experimental demonstration of the solenoid pion-capture and curved-solenoid muon-transport systems described above~\citep{Cook:2016sfz}. Construction of the MuSIC facility began in 2009, and measurements performed in 2011 demonstrated the high muon-production efficiency achieved with this configuration.

\begin{figure}[tb]
\centering
	 \includegraphics[width=0.6\textwidth]{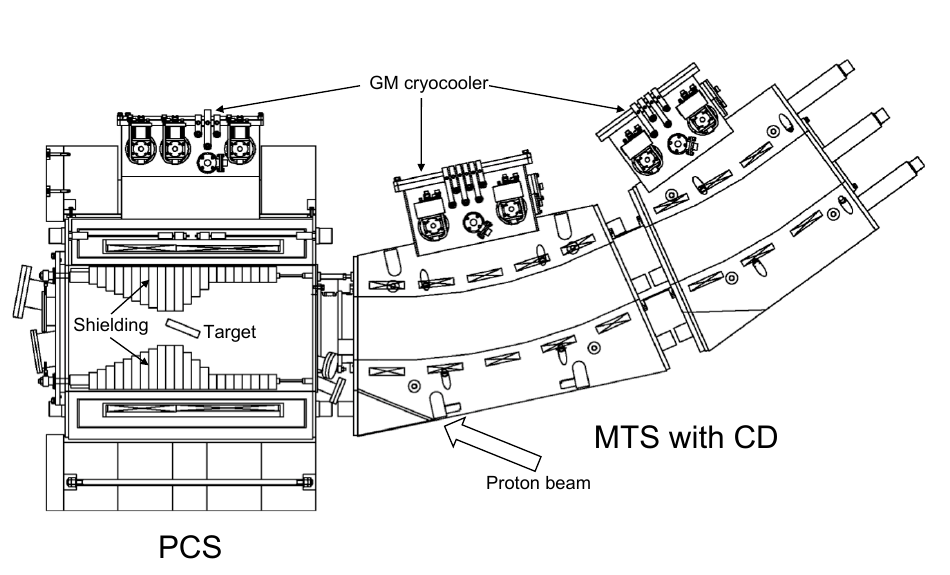}
\caption{Layout of the MuSIC muon source at the Research Center for Nuclear Physics (RCNP), The University of Osaka. The proton beam enters from the left and strikes the pion-production target located inside the pion-capture solenoid (PCS). The downstream curved muon-transport solenoid (MTS) incorporates correction dipoles (CDs) to control the charge- and momentum-dependent drift of the transported particles. Reproduced from \citet{Cook:2016sfz}, licensed under CC BY 4.0.}
\label{fig:music}
\end{figure}

The schematic layout of MuSIC is shown in \Cref{fig:music}, and a photograph of MuSIC is shown in \Cref{fig:music-photo}. The beamline consists of a superconducting pion-capture solenoid (PCS) followed by a superconducting curved muon-transport solenoid (MTS). Correction dipole magnets (CDs), installed in the MTS, generate a magnetic field perpendicular to the plane of curvature and are used to control the charge- and momentum-dependent transverse drift of the transported particles.

A $392\,\mathrm{MeV}$ proton beam from the RCNP ring cyclotron impinges on a cylindrical graphite pion-production target located inside the PCS. Backward-going pions are captured by the PCS, which provides a peak magnetic field of $3.5\,\mathrm{T}$, and are guided through a graded-field region into the MTS, which operates at $2\,\mathrm{T}$ and covers a bending angle of approximately $36^{\circ}$. The correction dipoles provide magnetic fields of up to $\pm 0.04\,\mathrm{T}$. The superconducting magnets use copper-stabilized NbTi conductors and are conduction-cooled by cryocoolers. MuSIC thereby also demonstrated that such a superconducting magnet system can operate in the radiation environment near a pion-production target and a proton beam dump.

By changing the polarity and magnitude of the correction dipole field, MuSIC demonstrated the preferential transport of positive or negative muons with selected momenta. Because only the first $36^{\circ}$ section of the curved transport system was commissioned, full charge and momentum separation was not expected. Nevertheless, the measurements established the practical operation of solenoid pion capture and curved-solenoid muon transport.

The negative-muon yield was measured using muonic X-rays emitted after the transported muons stopped in a target and was found to be of order $10^{5}\,\mu^-\,\mathrm{s}^{-1}\,\mathrm{W}^{-1}$. A separate measurement based on muon-decay spectra gave a combined positive- and negative-muon yield of order $10^{6}\,\mu^{\pm}\,\mathrm{s}^{-1}\,\mathrm{W}^{-1}$~\citep{Cook:2016sfz}. At the MuSIC proton energy of $392\,\mathrm{MeV}$, the positive-muon yield is much larger than the negative-muon yield, as discussed in \Cref{sec:principle}. These measurements demonstrated that solenoid pion capture and curved-solenoid muon transport can provide muon yields, normalized by proton-beam power, that are substantially higher than those of conventional muon beamlines. They thereby established that this concept can be applied to high-intensity searches for CLFV, including \mue\ conversion.

\section{PRISM/PRIME and the FFA Phase-Rotation Ring}
\label{app:prism-ffa}

As discussed in \Cref{sec:prism}, the PRISM concept uses a muon storage ring upstream of the muon stopping target to improve the purity and reduce the momentum spread of the muon beam. This appendix provides supplementary accelerator information, focusing on the choice of an FFA ring, the reference PRISM/PRIME accelerator concept, and the prototype demonstration of phase rotation.

\subsection{Motivation for an FFA Ring}
\label{app:prism-ffa-motivation}

A fixed-field alternating-gradient (FFA) ring\footnote{Such accelerators are abbreviated as FFAs and have historically also been referred to as FFAG accelerators.} is a circular accelerator in which static magnetic fields with designed field gradients confine and transport particles over a wide range of momenta. In such rings, focusing and defocusing magnetic-field elements are arranged alternately around the ring, providing stable transverse focusing of the beam. 
Because the magnetic field is static, particles with different momenta circulate on distinct, stable closed orbits~\citep{Symon:1956pr,Machida:2013irq}. For muon phase rotation, an FFA ring is suitable for the reasons described below.

The storage ring for muon phase rotation must accept a muon beam produced by solenoid pion capture. Such a beam has a large transverse emittance and a broad momentum spread, while the required longitudinal phase-space manipulation must be completed within a few turns because of the finite muon lifetime. These requirements favor an accelerator with large transverse and momentum acceptances, fixed-field operation, strong focusing, and the capability to perform RF phase rotation.

A conventional isochronous cyclotron provides fixed-field operation and relatively large momentum acceptance, but its nearly momentum-independent revolution period is poorly suited to RF phase rotation. A synchrotron can provide the required RF phase rotation, but its limited acceptance and the need for rapid magnetic-field ramping make it unattractive for a muon beam with large emittance. An FFA ring combines fixed-field operation with alternating-gradient focusing, allowing large beam acceptance and RF manipulation without magnetic-field ramping. For these reasons, an FFA ring was adopted as the baseline accelerator approach for PRISM.

For completeness, an alternative approach to residual pion suppression would be to use a long solenoidal decay channel, of order $100\,\mathrm{m}$, to provide an extended flight path before the muon stopping target.\footnote{In principle, RF cavities could also be installed along such a solenoidal channel to perform phase rotation, as in front-end concepts for a muon collider. However, this would require many RF cavities distributed over a long beamline.} In contrast, an FFA ring provides both residual pion suppression and RF phase rotation in a compact storage-ring geometry, in which the muon bunch repeatedly passes through the same RF system.

\subsection{Reference PRISM/PRIME Accelerator Concept}
\label{app:prism-prime-initial-study}

The original PRISM-FFA design employed a radial-sector scaling FFA lattice with defocusing--focusing--defocusing (DFD) triplet cells~\citep{Machida:2004hd,Sato:2006zza}. In such a lattice, the median-plane field can be written as $B(r)=B_0\left({r}/{r_0}\right)^k$, where $r_0$ and $B_0$ are reference values and $k$ is the field index. In the original design, $k=4.6$. The time-independent magnetic field allows particles with different momenta to circulate on different closed orbits, thereby accommodating the broad momentum spread of the injected muon beam.

Subsequent accelerator studies identified injection and extraction as major technical challenges, mainly because of the large emittance and orbit excursion of the muon beam. Advanced FFA concepts, including racetrack-like lattices with longer straight sections, were therefore investigated to facilitate the integration of the injection, extraction, and RF systems~\citep{Witte:2012zza,alekou2013acceleratorprismbasedmuon}. These studies demonstrated the potential of the PRISM/PRIME approach while leaving the detailed accelerator configuration as a subject for further design work.

\begin{figure}[tbh!]
\centering
	 \includegraphics[width=0.7\textwidth]{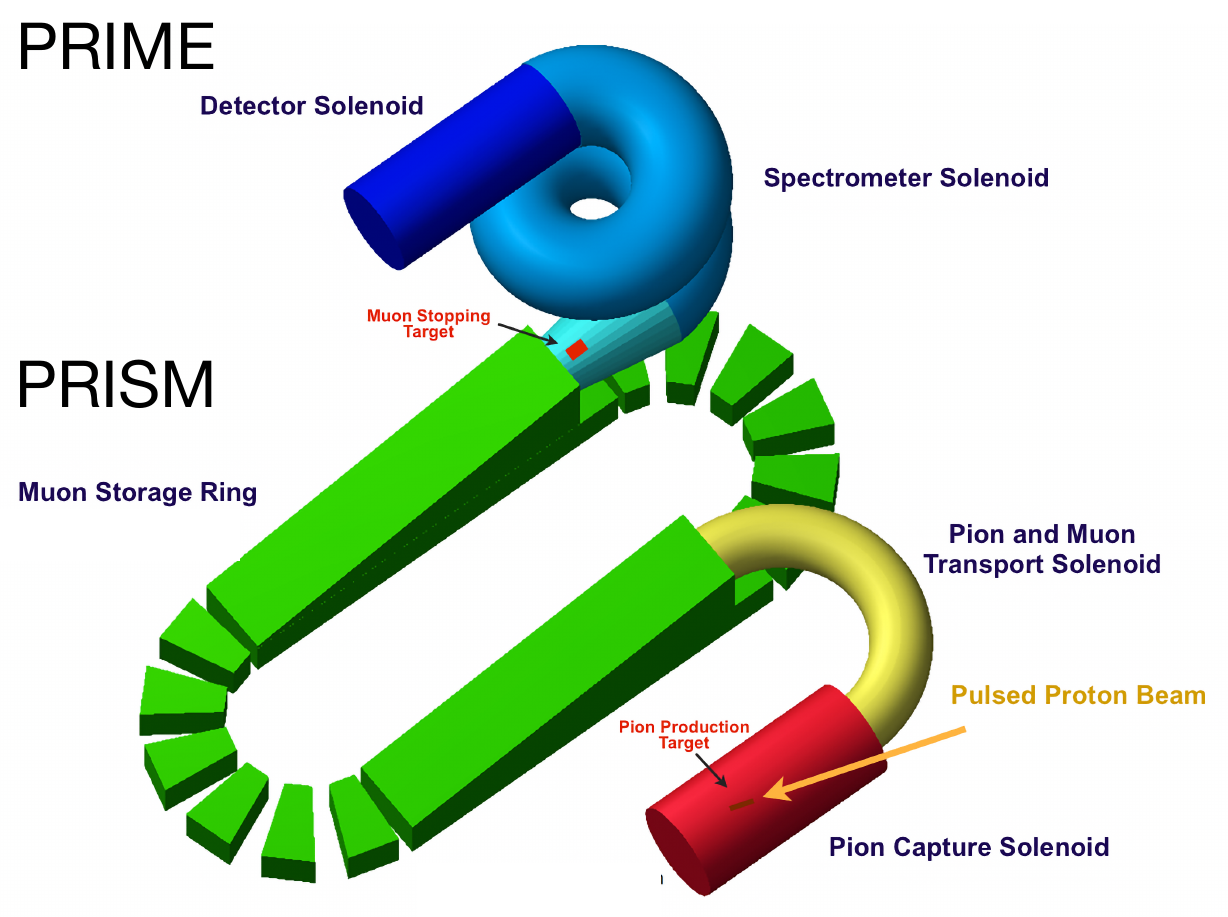}
\caption{Conceptual layout of the PRISM/PRIME accelerator and experimental system. Figure courtesy of the PRISM Collaboration.}
\label{fig:prism}
\end{figure}

A conceptual layout of the PRISM/PRIME accelerator and experimental system is shown in \Cref{fig:prism}. The layout includes a pion-capture solenoid, a transport solenoid, an FFA muon storage ring with a racetrack configuration for phase rotation, and a spectrometer solenoid providing more than $360^{\circ}$ of bending before the \mue\ conversion electrons reach the PRIME detector system. This layout should be regarded as a reference accelerator concept rather than a finalized experimental design.

\subsection{Prototype Demonstration of Phase Rotation}
\label{app:prism-ffa-prototype}

A prototype FFA ring was constructed at RCNP, The University of Osaka, to test the beam dynamics relevant to the PRISM phase-rotation scheme. The prototype ring was based on the original radial-sector scaling PRISM-FFA design and was assembled in late 2007 using six PRISM-FFA magnets and an RF cavity~\citep{Arimoto:2008zz}. A photograph of the prototype ring is shown in \Cref{fig:prototype-prism}. Each PRISM-FFA magnet cell employed a DFD-triplet lattice, and 
the diameter of the reference orbit was approximately $6.2\,\mathrm{m}$.

This configuration allowed the longitudinal phase-space dynamics of phase rotation to be studied without the need for a muon beam. The test used $\alpha$ particles from a ${}^{241}\mathrm{Am}$ source instead of muons. After passing through an aluminum degrader foil, the particles had momenta suitable for injection into and circulation in the ring. The reference circulating momentum was approximately $98\,\mathrm{MeV}/c$, corresponding to a revolution period of about $2.4\,\muup\mathrm{s}$.

The RF system operated at a frequency of approximately $2.1\,\mathrm{MHz}$ with harmonic number $h=5$ and a peak gap voltage of about $33\,\mathrm{kV}$~\citep{Sato:2008zze}. The RF phase was chosen so that higher-momentum particles were decelerated, while lower-momentum particles were accelerated, thereby reducing the momentum spread through phase rotation.

The observed manipulation of the longitudinal phase space demonstrated the essential operating principle of the PRISM-FFA phase-rotation ring~\citep{Sato:2010wepe044}. This experiment thus provided an important proof of principle for the accelerator concept underlying PRISM.

\section{The DeeMe Experiment}
\label{app:deeme}

The DeeMe experiment searches for \mue\ conversion at the Materials and Life Science Experimental Facility (MLF) of J-PARC~\citep{Natori:2014yba}. It uses a pulsed $3\,\mathrm{GeV}$ proton beam from the J-PARC Rapid Cycling Synchrotron (RCS), which is delivered to the MLF production target with a $25\,\mathrm{Hz}$ double-pulse time structure.

DeeMe provides an experimental approach complementary to those of COMET and Mu2e. Whereas COMET and Mu2e employ dedicated solenoid pion-capture and curved-solenoid muon-transport systems, DeeMe uses negative muons that stop in the pion-production target itself. The production target therefore also serves as the muon stopping target. In this scheme, some of the negative pions produced by the incident proton beam decay in flight before they can be absorbed by nuclei in the production target, and some of the resulting negative muons stop in the target and form muonic atoms. For the graphite target, Monte Carlo simulations for DeeMe estimate the probability of forming a muonic carbon atom to be of order $10^{-5}$ per incident $3\,\mathrm{GeV}$ proton~\citep{Nagao:2024zon}. The \mue\ conversion electrons emitted from the target are transported through the H beamline at the MLF to a magnetic spectrometer, where their momenta are measured.

This design makes DeeMe more compact than experiments that use separate pion-production and muon-stopping targets. However, since it also eliminates the spatial separation between the pion-production and muon-stopping regions, beam-related backgrounds associated with the pulsed beam structure and timing must be carefully controlled. In particular, the intense, short proton bunches of the high-power $25\,\mathrm{Hz}$ beam produce a large burst of prompt secondary particles when they strike the target. This beam flash results in extremely high instantaneous hit rates in the tracking detectors. The detectors must therefore tolerate these high rates and recover rapidly before the delayed measurement window opens.

With the present graphite production target at the MLF, DeeMe is expected to reach a single-event sensitivity (SES) of approximately $1\times10^{-13}$ after $2\times10^{7}\,\mathrm{s}$ of data taking at an RCS beam power of $1\,\mathrm{MW}$~\citep{Teshima:2019orf}. 
This sensitivity would improve on existing limits by up to about an order of magnitude. A possible silicon-carbide production target has also been discussed and could improve the projected SES by a factor of approximately five. Any such target change would, however, need to be considered in the context of the shared muon-production target used by other MLF muon beamlines.

DeeMe has a more modest projected sensitivity than COMET and Mu2e, but it provides a complementary search using a distinct and comparatively compact experimental concept.

%

\bibliographystyle{Harvard_noand}
\bibliography{references-kuno}

\typeout{get arXiv to do 4 passes: Label(s) may have changed. Rerun}

\end{document}